\documentclass[aps, pra,twocolumn,showpacs,superscriptaddress,floatfix]{revtex4-1} 
\usepackage{graphicx}
\usepackage{mathtools,amssymb,amsmath, amsthm,stmaryrd,thmtools,braket}
\usepackage{xr-hyper}
\usepackage[colorlinks=true, urlcolor=blue,citecolor=blue,anchorcolor=blue]{hyperref}
\usepackage[capitalise]{cleveref}

\usepackage{array}
\usepackage{bm}
\newcommand{\expect}{{\rm I\kern-.3em E}}
\newcommand{\Var}{\mathrm{Var}\,}

\newcommand{\norm}[1]{||#1||}
\newcommand{\abs}[1]{\left|#1\right|}
\let\vec\bm 
\newcommand{\vt}{\vec\theta}
\usepackage{hhline}
\usepackage{tikz}
\usetikzlibrary{arrows,shadows,positioning,shapes,calc,fadings,decorations.pathreplacing}

\usepackage{bbold}

\begin{document}
\title{Protocols for estimating multiple functions with quantum sensor networks: geometry and performance}
\author{Jacob Bringewatt}
\affiliation{Joint Center for Quantum Information and Computer Science, NIST/University of Maryland College Park, Maryland 20742, USA}
\affiliation{Joint Quantum Institute, NIST/University of Maryland College Park, Maryland 20742, USA}

\author{Igor Boettcher}
\affiliation{Department of Physics, University of Alberta, Edmonton, Alberta T6G 2E1, Canada}
\affiliation{Theoretical Physics Institute, University of Alberta, Edmonton, Alberta T6G 2E1, Canada}

\author{Pradeep Niroula}
\affiliation{Joint Center for Quantum Information and Computer Science, NIST/University of Maryland College Park, Maryland 20742, USA}
\affiliation{Joint Quantum Institute, NIST/University of Maryland College Park, Maryland 20742, USA}

\author{Przemyslaw Bienias}
\affiliation{Joint Center for Quantum Information and Computer Science, NIST/University of Maryland College Park, Maryland 20742, USA}
\affiliation{Joint Quantum Institute, NIST/University of Maryland College Park, Maryland 20742, USA}

\author{Alexey V. Gorshkov}
\affiliation{Joint Center for Quantum Information and Computer Science, NIST/University of Maryland College Park, Maryland 20742, USA}
\affiliation{Joint Quantum Institute, NIST/University of Maryland College Park, Maryland 20742, USA}

\date{\today}

\begin{abstract}
    We consider the problem of estimating multiple analytic functions of a set of local parameters via qubit sensors in a quantum sensor network. To address this problem, we highlight a generalization of the sensor symmetric performance bounds of Rubio \textit{et al}
    [\textit{J. Phys. A: Math. Theor.} \textbf{53} 344001 (2020)] and develop a new optimized sequential protocol for measuring such functions. 
    We compare the performance of both approaches to one another and to local protocols that do not utilize quantum entanglement, emphasizing the geometric significance of the coefficient vectors of the measured functions in determining the best choice of measurement protocol. We show that, in many cases, especially for a large number of sensors, the optimized sequential protocol results in more accurate measurements than the other strategies. 
    In addition, in contrast to the the sensor symmetric approach, the sequential protocol is known to always be explicitly implementable. 
    The sequential protocol is very general and has a wide range of metrological applications. 
\end{abstract}
\maketitle

\section{Introduction}
It is well-established that entanglement in quantum metrology often facilitates more accurate measurements compared to what is possible with unentangled probes \cite{bollinger1996optimal, huelga1997improvement, pezze2009entanglement, toth2012multipartite,  zhang2014quantum}. This fact has been demonstrated exhaustively for the cases of measuring a single parameter \cite{Boixo2007} or a single analytic function of many parameters \cite{Eldredge2018, proctor2017networked, proctor2018multiparameter, altenburg2018multi, Qian2019, gross2020one, qian2020optimal, triggiani2021heisenberg} using quantum sensor networks, which are highly general models of quantum metrology. In these models, one considers an array of $d$ quantum sensors, each coupled to a local parameter. One then seeks to optimally measure these local parameters directly (or some functions thereof) by selecting an initial state $\rho_0$ for the sensors, a unitary evolution $U$ by which the local parameters are encoded in the state, and a choice of measurement specified by a positive operator-valued measure (POVM). 

While measuring a single analytic function of multiple parameters in this setting is a bona fide multi-parameter problem, the fact that one seeks a single quantity makes the problem of finding the information-theoretic optimum for the variance of the desired quantity easier than a more general multi-parameter problem; in particular, one can make clever use of rigorous bounds originally derived for the single-parameter case \cite{Eldredge2018,Qian2019,gross2020one}. However, when one genuinely seeks to estimate multiple quantities, one must solve the general problem of designing provably optimal protocols for multi-parameter quantum estimation. This has proven to be a challenging problem, and has attracted a large amount of interest theoretically \cite{helstrom1976quantum, holevo2011probabilistic, paris2009quantum, genoni2013optimal, zhang2014quantum, zhang2014quantum2, yue2014quantum, gao2014bounds, knott2016local, baumgratz2016quantum, ragy2016compatibility, szczykulska2016multiparameter, pezze2017optimal, hall2018entropic, gessner2018sensitivity, altenburg2018multi, zhuang2018distributed, yang2019attaining, albarelli2019evaluating, tsang2019holevo, albarelli2019upper, sidhu2020geometric, rubio2020bayesian, rubio2020quantum} and experimentally \cite{vidrighin2014joint, roccia2018multiparameter, valeri2020experimental}. Despite these extensive research efforts, the general problem has not yet been solved. Here, we consider another step towards this goal; in particular, we consider the case of measuring $n\leq d$ analytic functions with a quantum sensor network of $d$ qubit sensors and develop a protocol that outperforms previously proposed protocols in many cases. We also emphasize the geometric aspects of this problem, meaning the orientations of vectors of coefficients associated with our functions, and how this geometry determines the protocol performance. 

We begin by noting that, analogous to Ref. \cite{Qian2019}, one can reduce the problem of measuring $n$ analytic functions of the parameters to that of measuring $n$ linear functions. In particular, one can consider spending some asymptotically (in total time $t$) vanishing time $t_1$ measuring the local parameters to which the sensors are coupled and then the rest of the time $t_2=t-t_1$ measuring the $n$ linear combinations that result from a Taylor expansion of each analytic function about the true values of the local parameters estimated in the previous step. While provably optimal in the single-function case ($n=1$), this reduction from analytic functions to linear functions is not necessarily optimal in the multi-function case. While we conjecture that the optimality of this reduction from analytic to linear functions \emph{does} generalize to the multi-function case, as we do not claim general optimality of the protocols in this work, the reduction may be freely made without having to prove the veracity of this conjecture.

\begin{figure*}
    \centering
    \includegraphics[width = 0.85\textwidth]{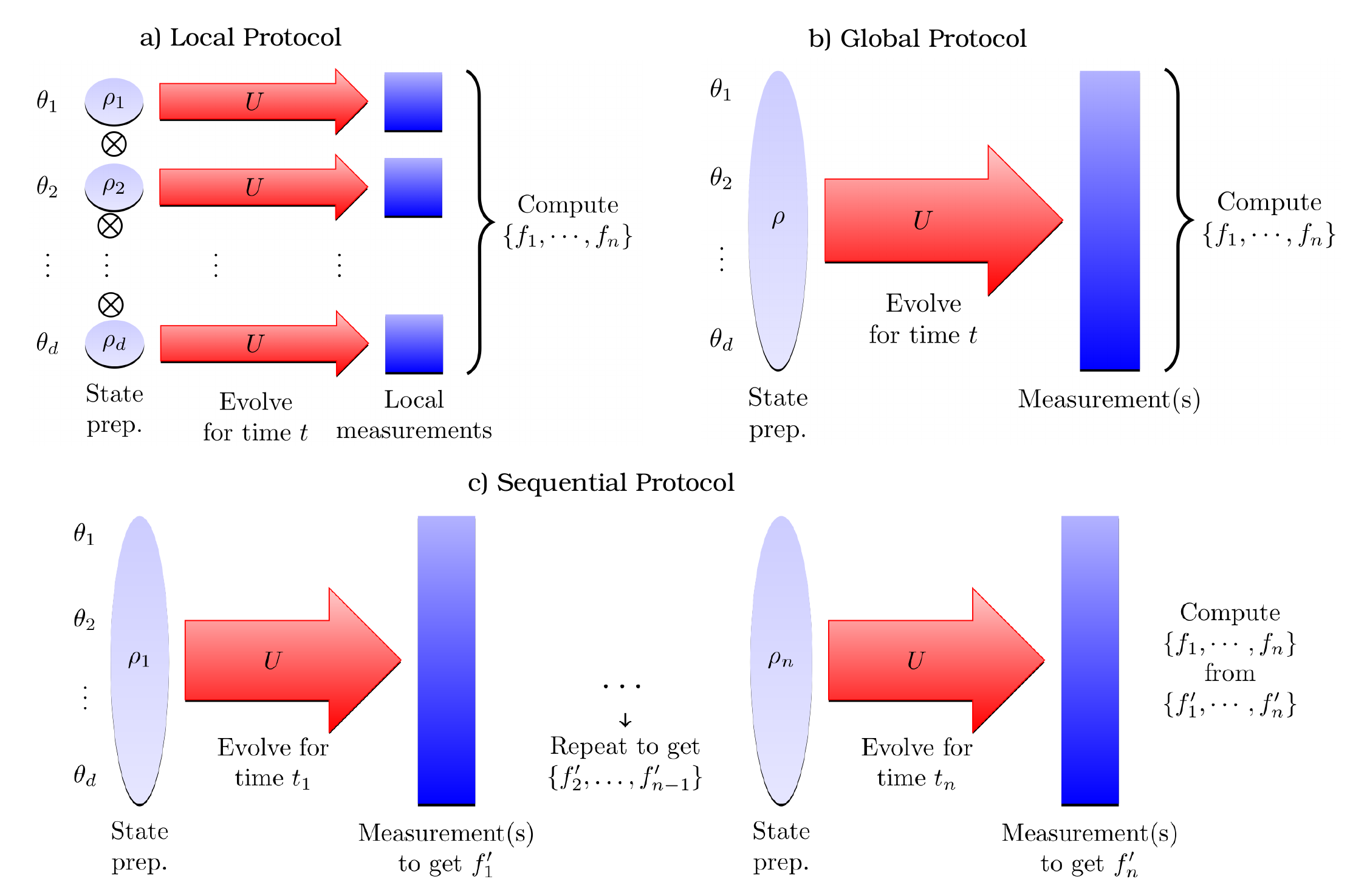}
    \caption{The protocols for measuring $n\leq d$ linear functions $\{f_1(\vec{\theta}),\dots,f_n(\vec{\theta})\}$ of $d$ parameters $\vec{\theta}=(\theta_1,\dots,\theta_d)$ considered in this work can be classified into three groups: (a) Local protocols do not utilize entanglement and measure the parameters locally, allowing for large parallelization. (b) Global protocols simultaneously estimate all functions. (c) Sequential protocols divide the problem into $n$ parts, where each part is optimized to estimate a single function from the set $\{f_1',\dots,f_n'\}$, which may consist of linear combinations of the original set $\{f_1,\dots,f_n\}$.}
    \label{fig:protocoltypes}
\end{figure*}

Having made this reduction to the problem of measuring multiple \emph{linear} functions in a quantum sensor network, we can connect to previous works addressing the same problem, subject to various simplifying constraints \cite{proctor2017networked, altenburg2018multi, rubio2020quantum}.      
Leaving the details of these previous approaches for after we have introduced more mathematical formalism, we note that we may qualitatively divide protocols for this problem into three classes: local, global, and sequential \cite{altenburg2018multi}. In a local estimation protocol, one optimizes only over unentangled input states and local measurements of the sensors. In a global protocol, one simultaneously estimates all the desired functions by optimizing over all (possibly entangled) input states and all (possibly non-local) measurements. Finally, in a sequential protocol, we divide the experiment into $n$ steps, where in each part we measure a single function (which may be a linear combination of the original set $\{f_1, \cdots, f_n\}$),
preparing a new (optimal) initial state and performing a new measurement in each step.  See \cref{fig:protocoltypes} for diagrammatic representations of these different protocol types.

For the special case of measuring $n=d$ orthogonal, linear functions, it has been known for some time that the functions can be measured optimally with a local protocol \cite{proctor2017networked, altenburg2018multi}, but for general functions, proofs of optimal protocols are lacking. In fact, the only entanglement-enhanced approach in the literature for measuring $n>1$ general linear functions in a quantum sensor network is given in Ref.~\cite{rubio2020quantum}. The bound on performance given there is for global protocols and is derived from the quantum Cram\'{e}r-Rao bound \cite{holevo2011probabilistic, helstrom1976quantum, braunstein1994statistical, braunstein1996generalized} subject to the restriction that one considers only a special set of so-called sensor symmetric states. However, even within this restriction, beyond the case of $d=2$, it is an open question whether the states and measurements (POVMs) required to saturate the derived bound exist for all problems \footnote{The reason that there may not exist states satisfying the bound is that, as explained later in the paper, the bound is obtained by fixing $v$ [defined in Eq.~(\ref{eq:ssstates})] to be $t^2/4$ and then optimizing $\mathcal{J}$ [defined in Eq.~(\ref{eq:intersensorcorrelations})] given this restriction to obtain the best bound. In principle, we are not guaranteed a state corresponding to this pair of $4v=t^2$ and the minimizing $\mathcal{J}$, but of course the bound is still a correct lower bound if one is only allowed to use sensor symmetric states whether or not it can be saturated. Also, see Ref.~\cite{rubio2020quantum} for further discussion.}.

Here, we highlight a generalization of this approach, by deriving similar bounds using so-called \emph{signed sensor symmetric states}. However, the generalized version also does not guarantee that the optimal states and measurements exist in general. Targeted at this shortcoming, we also consider an alternative, sequential protocol, subject to different restrictions, for which we can explicitly describe a protocol which achieves its theoretical performance. In addition to presenting this alternative protocol, we lay out how the precise geometric features of a given problem impact the performance of this sequential protocol compared to the signed sensor symmetric approach and the simple local protocol.

\section{Problem Setup}
With the general approach established, we now present the rigorous formulation of the problem. We consider a quantum sensor network of $d$ qubit sensors prepared in some initial state $\rho_0$. We then encode $d$ local parameters $\vec{\theta}=(\theta_1, \theta_2, \cdots, \theta_d)^T$ into the sensors via unitary evolution under the Hamiltonian
\begin{equation}\label{eq:H}
    \hat{H}=\hat{H}_{\rm c}(t)+\sum_{i=1}^d\frac{1}{2}\theta_i\hat{\sigma}_i^{z},
\end{equation}
with $\hat{\sigma}_i^{x,y,z}$ the Pauli operators acting on the $i^\mathrm{th}$ qubit,
and $\theta_i$ the local parameter measured by the $i^{\mathrm{th}}$ sensor. The term  $\hat{H}_{\rm c}(t)$ is a time-dependent control Hamiltonian that may include coupling to ancilla qubits. When measuring a single function, this time-dependent control is not necessary to achieve an optimal protocol \cite{Boixo2007, Eldredge2018}, but one may use such control to design optimal protocols with simpler requirements on the choice of input state $\rho_0$ \cite{Eldredge2018}. 
Using this setup, our goal is to optimally measure $n\leq d$ functions $\vec f(\vec{\theta})=(f_1(\vt), f_2(\vt), \cdots, f_n(\vt))^T$.  In the following, we use $i,j=1,\dots, d$ to label qubits and $\ell, m=1,\dots,n$ to label functions. Boldface is used to denote vectors.

To compare the accuracy of the different approaches and to eventually optimize them, we employ a standard figure of merit, which we denote as $\mathcal{M}$, given as
\begin{equation}
  \label{EqDeltaE}  \mathcal{M}=\sum_{\ell=1}^n w_\ell \Var{\tilde{f}_\ell},
\end{equation}
where $\vec{\tilde{f}}$ are estimators of the functions and $\vec w=(w_1,\cdots, w_n)^T$ is a vector of weights.
Since an accurate protocol should yield small variances, we seek to minimize $\mathcal{M}$. 
In this context, given a total evolution time $t$, a protocol is defined by choice of initial state $\rho_0$, control Hamiltonian $\hat{H}_{\rm c}(t)$, measurements, and estimator $\vec{\tilde f}$ for $\vec f$. 

The figure of merit $\mathcal{M}$ is lower bounded via  the Helstrom quantum Cram\'{e}r--Rao bound \cite{holevo2011probabilistic, helstrom1976quantum, braunstein1994statistical, braunstein1996generalized}, which yields
\begin{equation}\label{eq:cramerrao}
    \mathcal{M}\geq \frac{1}{N}\sum_{\ell=1}^n w_\ell [\mathcal{F}_Q^{-1}(\vec f)]_{\ell\ell}, 
\end{equation}
where $N$ is the number of trials (which from now on we set to one for concision and consider just the single-shot Fisher information) and $\mathcal{F}_Q(\vec f)$ is the quantum Fisher information matrix with respect to the functions $\vec f$. While this bound is not generally saturable, in the setting of Eq.~(\ref{eq:H}) it is \footnote{In particular, it is saturable because the generators of translations $K_i$, as defined in the discussion around Eq.~(\ref{eq:intersensorcorrelations}), commute.}.

While saturable in the setting considered, the right hand side of \cref{eq:cramerrao} is not easily evaluated in general. However, it has been proven \cite{Eldredge2018} that, if we seek to measure a \emph{single} linear function $f(\vt)=\vec{\alpha}\cdot\vec{\theta}$ of the parameters $\vt$, we may evaluate this bound and obtain that the minimum (asymptotically in time $t$ and number of trials) attainable variance of an estimator $\tilde f$ of $f(\vt)$ over all quantum protocols is 
\begin{equation}
  \label{EqEld}  \Var{\tilde f} = \max_i\frac{|\alpha_i|^2}{t^2}.
\end{equation}
This bound can be explicitly saturated by the protocols given in Ref. \cite{Eldredge2018}. As previously described, if $f(\vec{\theta})$ is a more general analytic function, one may attain a similar bound using a two-step protocol. In the first (asymptotically negligible) step, one makes local estimates $\vec{\tilde\theta}$ of each of the parameters $\vec\theta$. In the second step, one uses the rest of the time to optimally measure the Taylor expansion of $f(\vt)$ about this estimate to linear order in $\vec\theta$ \cite{Qian2019}. 

For the case of measuring multiple functions $f_1,\dots,f_n$, we assume without loss of generality that the $f_\ell$ are \emph{linear functions} in the parameters $\vec{\theta}$, because more general analytic functions could be similarly linearized in asymptotically negligible time. We parameterize the linear functions by real \emph{coefficient vectors} $\vec{\alpha}_\ell$ such that
\begin{align}
    f_1(\vec{\theta}) &= \vec{\alpha}_1\cdot \vec{\theta},\\
    \nonumber &\vdots\\
    f_n(\vec{\theta}) &= \vec{\alpha}_n\cdot \vec{\theta}.
\end{align}
Defining the matrix elements $A_{\ell i}=(\partial f_{\ell}/\partial \theta_i)_{\vec{\tilde{\theta}}}=(\vec{\alpha}_\ell)_i$, i.e., $\vec{\alpha}_\ell^T$ is the $\ell$th row of $A$, we can phrase the problem as that of optimally measuring the $n$-component vector
\begin{equation}
 A\vec\theta = \left( \vec{\alpha}_1\cdots\vec{\alpha}_n\right)^T\vec\theta.
\end{equation}
Without loss of generality we assume normalization of the coefficient vectors,
\begin{equation}\label{eq:normcond}
    ||\vec{\alpha}_\ell||^2=1\text{ for all }\ell,
\end{equation}
because any non-unit length can be absorbed into the  weights $\vec w$ in Eq.~(\ref{EqDeltaE}).

Recall, the problem of measuring $n=d$ linear functions of independent parameters with quantum sensor networks has been considered in the literature in the case where the $n$ functions are orthogonal (in which case local, global and sequential protocols are equivalent) \cite{proctor2017networked, altenburg2018multi} and for general linear functions for global protocols when the input states $\rho_0$ are restricted to be sensor symmetric \cite{rubio2020quantum}. Here, we generalize the sensor symmetric approach and derive a performance bound when using so-called signed sensor symmetric input states (defined rigorously below). We refer to the variance obtained by the signed sensor symmetric protocol as $\mathcal{M}_{\mathrm{ss}}$. 

In this work, we also introduce an optimized sequential protocol for solving the $n$ function estimation problem. We consider dividing our protocol into $n$ sequential steps where, \emph{within each step}, the protocol is provably information-theoretic optimal (i.e.,~saturates the quantum Cram\'{e}r--Rao bound). In particular, for each step $\ell\in \{1,\ldots,n\}$ taking time $t_\ell$, we measure a single function optimally using the protocols from Refs. \cite{Eldredge2018, Qian2019}. We cannot, however, prove that the full protocol is optimal in an information-theoretic sense. The naive version of this protocol is to measure the $n$ given functions $\{f_1,\dots,f_n\}$ one after another with some optimal choice of the time $t_\ell$ spent on each function. We denote the figure of merit of the naive sequential protocol by $\mathcal{M}_\mathrm{naive}$. 

\begin{figure}
    \centering
   \includegraphics[width =\columnwidth]{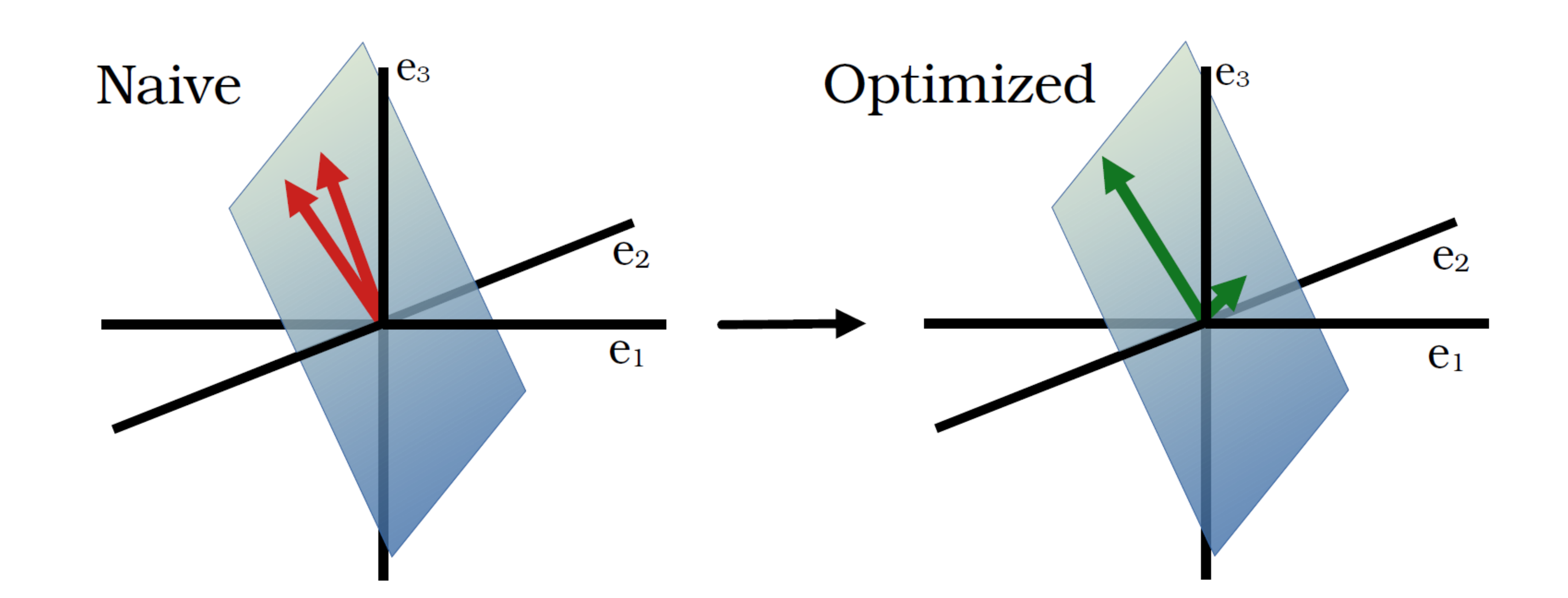}
    \caption{A visualization for $n=2$ functions and $d=3$ sensors of how we can optimally select a set of functions to measure whose coefficient vectors $\{\vec\alpha_\ell'\}$ span the same subspace as the coefficient vectors $\{\vec\alpha_\ell\}$ of the functions we care about. The vectors are the coefficient vectors and the planes indicate the subspace they span. The axes are labeled by standard basis unit vectors $\{e_1, e_2, e_3\}$. }
    \label{fig:visualizeopt}
\end{figure}

However, the naive sequential protocol is not the only option for sequentially measuring multiple functions. Indeed, the coefficient vectors $\{\vec{\alpha}_1, \cdots, \vec{\alpha}_n\}$ span a linear subspace of $\mathbb{R}^d$, and we may instead sequentially measure \emph{any} set of linear functions whose vectors of coefficients $\{\vec{\alpha}'_1, \cdots, \vec{\alpha}'_n\}$ span the same subspace and then (after the measurements) calculate the original functions $\{f_1,\dots,f_n\}$. To help understand this visually, this approach is depicted in the diagram in Fig.~\ref{fig:visualizeopt} for $n=2$ functions and $d=3$ sensors. We denote the figure of merit obtained via this method by $\mathcal{M}_\mathrm{opt}$. 

To be explicit, define the $n\times n$ matrix $C$ encoding the change of linear functions via 
\begin{align}
  \label{EqACA}  A=CA',
\end{align}
where $A'=\left( \vec{\alpha}'_1, \cdots, \vec{\alpha}'_n\right)^T$ is the matrix whose rows are the coefficient vectors of the new linear functions we measure. The variance of measuring any individual $\vec{\alpha}'_\ell$ is given by the optimal linear protocol \cite{Eldredge2018}
\begin{equation}
    \mathcal{M}_\ell=\frac{{\mu'_\ell}^2}{t_\ell^2},
\end{equation}
where we introduce
\begin{equation} \label{eq:mudef}
\mu'_\ell=\norm{\vec{\alpha}_\ell'}_\infty=\max_{j}\ |\alpha'_{\ell,j}|=\max_{j}\abs{\sum_{m=1}^n(C^{-1})_{\ell m}A_{m j}}.
\end{equation}
Note that this corresponds to Eq.~(\ref{EqEld}) for every $\ell$. We denote by $\vec{\mu}'$ the vector with entries $\mu'_\ell$, and by $\vec\mu$ the analogous vector for the original functions [obtained by setting $C=I$ in Eq.~(\ref{eq:mudef})].
The figure of merit for estimating the original functions $\vec{f}$ with the optimized sequential protocol is then formally given by
\begin{equation}\label{eq:eoptorig}
    \mathcal{M}_{\mathrm{opt}}=\min_C\min_{\{t_1,\cdots, t_n\}}\left[\sum_{\ell=1}^n\sum_{m=1}^n w_m C_{m \ell}^2\left(\frac{\mu'_\ell}{t_\ell}\right)^2 \right],
\end{equation}
which takes into account optimization over $C$ and over the division of the total time into time steps $t_\ell$;  the factor $C_{m \ell}^2$ comes from the standard expression for a linear combination of variances and accounts for the linear change of functions. A more practical form of $\mathcal{M}_{\rm opt}$ will be derived below.
If the naive sequential protocol were optimal, then the minimum of $\mathcal{M}_{\rm opt}$ would be attained at $C=I$. However, we will show in the following that choosing suitable $C\neq I$ often gives a significant improvement. This matches one's intuitive expectations --- for example, if the coefficient vectors of all the functions are nearly aligned, we might expect that the optimal approach is to spend most of the time measuring a single function whose coefficient vector is in that general direction, and the rest of the time measuring functions with orthogonal coefficient vectors to distinguish the small differences in the functions we care about. We will see that this intuition is correct.

Furthermore, we note that for this approach, we do not consider taking advantage of potential parallelization that may arise for certain choices of functions to measure---in particular, those sets of functions that depend on completely disjoint sets of sensors. More formally, when one chooses functions to measure such that $A'$ is the direct sum of matrices representing linear functions on disjoint sets of qubits, one could simultaneously measure functions that depend on disjoint sets of sensors, and thus spend more time measuring them, improving the accuracy. Therefore, purposefully choosing functions to measure that allow for such parallelization could potentially (although not necessarily) perform better than our protocol, which does not take this possibility into account. However, improved performance via parallelization is not guaranteed as Eq.~(\ref{eq:eoptorig}) depends on both the time $t_\ell$ spent measuring a function \emph{and} the infinity-norm of the coefficient vector, $\mu_\ell'=\norm{\vec{\alpha}_\ell'}_\infty$---whereas parallelization improves the former, it may worsen that latter. 

We note that, when $n=d$, the local strategy is a special case of such parallelization as it consists of measuring the local parameters all in parallel, and therefore a completely diagonal $A'$.  As another simple example, suppose $\vec\alpha_1=(1,1,1)^T/\sqrt{3}$, $\vec\alpha_2=(1,-1,1)^T/\sqrt{3}$, and $\vec\alpha_3=(0,0,1)^T$. One way (amongst several) that this could be parallelized would be choosing to measure $\vec\alpha_1'=(1,1,0)^T/\sqrt{2}$, $\vec\alpha_2'=(1,-1,0)^T/\sqrt{2}$, and $\vec\alpha_3'=(0,0,1)^T$; with this choice, one could, in parallel, estimate the sets of functions $\{\vec\alpha_1', \vec\alpha_2'\}$ and $\{\vec\alpha_3'\}$.

At this point, we have commented on four approaches to our problem: (1) the local strategy with variance $\mathcal{M}_\mathrm{local}$ (defined in Eq.~(\ref{eq:elocal}) below), (2) the (global) signed sensor symmetric strategy generalized from Ref.~\cite{rubio2020quantum} with variance $\mathcal{M}_\mathrm{ss}$, (3) the naive sequential strategy with variance $\mathcal{M}_\mathrm{naive}$, (4) the optimized sequential strategy with variance $\mathcal{M}_\mathrm{opt}$. Importantly, \emph{none} of these strategies is optimal in general. Depending on the geometry of the linear functions to be measured, each of these strategies could be the preferable one (excluding the naive strategy, which, of course, in the best case, has $\mathcal{M}_\mathrm{naive}=\mathcal{M}_\mathrm{opt}$). The term ``geometry'' here refers to the absolute and relative orientations of the coefficient vectors $\{\vec\alpha_\ell\}$. The question of what is the ultimate information-theoretic limit on $\mathcal{M}$ for multiple linear functions remains open. Here, we demonstrate cases in which each of these known strategies is preferable with an emphasis on the geometric interpretation. We emphasize that, in many instances, both the signed sensor symmetric and the optimized sequential strategy can out-perfom the local unentangled strategy, which is of great importance for practical applications.

\section{The Strategies} \label{s:protocols} 
In this section, we determine the figure of merit $\mathcal{M}$ for the four strategies considered in this work. We emphasize that while the local and sequential strategies have explicit protocols to obtain the corresponding figure of merit, the figure of merit for the signed sensor symmetric is not proven to be always be attainable beyond $d=2$.

\subsection{Local Strategy}
First we consider the local strategy, which does not utilize entanglement. Since we can measure each local parameter $\theta_i$ simultaneously, with a variance of $1/t^2$ \cite{wineland1992spin}, we arrive at
\begin{align}\label{eq:elocal}
    \mathcal{M}_\mathrm{local}=\sum_{\ell=1}^n w_\ell \frac{||\vec{\alpha}_\ell||^2}{t^2}
    =\frac{1}{t^2}\sum_{\ell=1}^n w_\ell
    =\frac{\mathcal{N}}{t^2},
\end{align}
where we used the normalization of the $\vec{\alpha}_\ell$ and introduce
\begin{equation}\label{eq:norm}
    \mathcal{N}=\sum_{\ell=1}^n w_\ell.
\end{equation}
We emphasize that the local protocol performs independently of the geometry of the measured linear functions.

\subsection{Signed Sensor Symmetric Strategy}
Next we review the results of Ref.~\cite{rubio2020quantum} for the sensor symmetric approach, using our notation and emphasize a generalization of their approach to what we call signed sensor symmetric states. We emphasize that, given the restriction to (signed) sensor symmetric states, this approach gives a rigorous lower bound on the figure of merit $\mathcal{M}$. However, as previously discussed, unlike the local or sequential strategies, for $d>2$ one cannot guarantee that the figure of merit $\mathcal{M}_\mathrm{ss}$ obtained via this approach is saturable \cite{rubio2020quantum}. 

Define the generators of translations in parameter space as $\vec K=(K_1,\cdots,K_d)^T$, where $K_i=\text{i}(\partial U/\partial \theta_i)U^\dagger$ for evolution under the unitary $U$. Following Ref.~\cite{rubio2020quantum}, for this strategy, we specifically consider the Hamiltonian in Eq.~(\ref{eq:H}) with $\hat{H}_c(t)=0$, so that $U=\exp(-i\hat{H}t)$ and $K_i=\hat{\sigma}_i^{z} t/2$. This restriction of \cref{eq:H} to evolution under a time-independent Hamiltonian is not necessary for the sequential protocols considered later. However, the single linear function results from Ref.~\cite{Eldredge2018}, which we use as a subroutine of our sequential protocol, presents two protocols, one that matches this restriction and one that does not (see section IV therein). Therefore, when explicitly comparing the sequential protocol to the signed sensor symmetric problem, we assume we are considering the former. 

Given the generators of translations $K_i$, we define the inter-sensor correlations \cite{knott2016local, proctor2017networked} by
\begin{equation}\label{eq:intersensorcorrelations}
    \mathcal{J}_{ij}=\frac{\langle K_iK_j\rangle - \langle K_i \rangle \langle K_j \rangle }{\Delta K_i \Delta K_j}
\end{equation}
for $i\neq j$, where we have used $\left(\Delta K_i\right)^2=\langle K_i^2\rangle - \langle K_i \rangle^2$. Given this definition, we define \emph{sensor symmetric states} as those such that for all $i\neq j$,  $\mathcal{J}_{ij}=\mathcal{J}=c/v$ with
\begin{align}\label{eq:ssstates}
    v=\langle K_i^2\rangle - \langle K_i \rangle^2, & & c=\langle K_iK_j\rangle - \langle K_i \rangle\langle K_j \rangle.
\end{align}
Specifically, for evolution under the time-independent version of \cref{eq:H}. we have
\begin{align}\label{eq:vc}
    v=\frac{t^2}{4}\left(1 - \langle \sigma^{z}_i \rangle^2\right) && c=\frac{t^2}{4}\left(\langle \sigma^{z}_i\sigma^{z}_j\rangle - \langle \sigma^{z}_i \rangle\langle \sigma^{z}_j \rangle\right)
\end{align}
for all $i\neq j$. The authors of Ref.~\cite{rubio2020quantum} define such states in analogy with path-independent states in optical interferometry \cite{hofmann2009all, knott2016local}, which, in addition to the analytic accessibility provided by such states, motivates this construction. The case of uncorrelated sensors, of course, is included for $\mathcal{J}=0$. 

Now we turn to a generalization of the sensor symmetric states considered in Ref.~\cite{rubio2020quantum} that we call signed sensor symmetric states. This generalization is natural as the (unsigned) sensor symmetric state construction of Ref.~\cite{rubio2020quantum} picks out functions with coefficient vectors $\vec\alpha$ aligned along the vector of all ones $\vec{1}=(1,1,\cdots,1)^T$ as being favorable, but we know the positive orthant is not special, and one can immediately generalize from $\vec{1}$ being the favorable orientation to any $\vec\omega\in\{-1,1\}^d$ (of which $\vec{1}$ is just one example). The reason such functions are most favorable is also intuitively clear---entanglement is most helpful when one measures global, average-like quantities, which is precisely what functions with coefficient vectors aligned along some $\vec\omega$ are. We emphasize this generalization is very direct, as one can consider mapping any problem using a general $\vec\omega$ to the case of Ref.~\cite{rubio2020quantum} merely by applying a Pauli-$X$ operator on all qubit sensors corresponding to negative elements of $\vec\omega$ and correspondingly flipping the signs of all corresponding coefficients specified by $\vec\alpha_\ell$. However, to fairly compare to the sequential protocol, it is important we consider all such $\vec\omega$, as different choices can lead to an improved figure of merit. Therefore, we relax the restriction on the numerator of $\mathcal{J}_{ij}$ as presented in Ref.~\cite{rubio2020quantum} by defining
\begin{align}
    c_{ij}=\langle K_iK_j\rangle - \langle K_i \rangle\langle K_j \rangle
\end{align}
and then restrict our consideration to states such that 
\begin{align} \label{eq:signedsensorstates}
    c_{ij}= c \left(\vec\omega\vec\omega^T\right)_{ij}=c\ \Omega_{ij},
\end{align}
where $\vec\omega\in\{-1,1\}^d$ is a vector with all entries $\pm 1$ and $c$ is a constant. The entries of $\Omega_{ij}$ are also $\pm 1$ and so $c_{ij}=\pm c$. We keep the definition $\mathcal{J}=c/v$ for our newly defined $c$, but note that now $\mathcal{J}_{ij}=c_{ij}/v=\pm\mathcal{J}$. 

When restricted to the (unsigned) sensor symmetric initial states, i.e. when $\vec{\omega}=\vec{1}$ with $\vec{1}=(1,\dots,1)^T$ the vector of all ones, the authors of Ref.~\cite{rubio2020quantum} were able to evaluate the quantum Cram\'{e}r--Rao bound and determine the minimal achievable value for $\mathcal{M}$, given the requirement of sensor symmetric input states. For the signed sensor symmetric states, the calculation is similar to that in Ref.~\cite{rubio2020quantum}, so we just state the result for our generalized approach here and present the details in Appendix \ref{app:signedsensorsymmetric}. 

First define the $\vec\omega$-dependent geometry parameter $\mathcal{G}(\vec\omega)$, which encodes the geometric relationship between the coefficient vectors $\{\vec{\alpha}_\ell\}$ of the $n$ linear functions and the vector $\vec\omega$. We have
\begin{equation}\label{eq:G}
    \mathcal{G}(\vec\omega)=\frac{1}{\mathcal{N}}\sum_{\ell=1}^n w_\ell\left(d\cos^2\phi_{\vec \omega, \ell}-1 \right).
\end{equation}
Here $\phi_{\vec \omega, \ell}$ is the angle between the vectors $\vec{\alpha}_\ell$ and $\vec \omega$. Thus $\cos \phi_{\vec \omega, \ell}=\vec{\alpha}_\ell\cdot\vec{\omega}/\sqrt{d}$. Note that $\mathcal{G}(\vec\omega)\in[-1, d-1]$. Again, we note that the relevance of this geometric quantity is intuitively clear as entanglement provides the biggest benefit when measuring functions aligned along some $\vec\omega$---that is, those functions for whom $\phi_{\vec \omega, \ell}\approx 0$. 
The $\vec\omega$-dependent lower bound on the figure of merit is found to be
\begin{equation}\label{eq:sscost}
    \mathcal{M}_{\mathrm{ss}}(\vec\omega)=\min_{\mathcal{J}} \frac{\mathcal{N}}{t^2}\frac{1+[d-2-\mathcal{G}(\vec\omega)]\mathcal{J}}{(1-\mathcal{J})[1+(d-1)\mathcal{J}]},
\end{equation}
where we have used $4v=t^2$ as in Ref.~\cite{rubio2020quantum} to obtain the lowest bound. Under this condition on $v$, and the assumption that $\mathcal{J}\in (1/(1-d), 1)$, so that the quantum Fisher information is invertible, the minimum is attained for 
\begin{align}\label{eq:Jopt}
   \mathcal{J}_\mathrm{opt}&(\vec\omega)=\nonumber \\
    &\frac{1}{\mathcal{G}(\vec\omega)+2-d}\left[ 1-\sqrt{\frac{(\mathcal{G}(\vec\omega)+1)[d-1-\mathcal{G}(\vec\omega)]}{d-1}}\right].
\end{align}
One can then obtain the theoretical best performance for a signed sensor symmetric strategy as
\begin{equation}
    \mathcal{M}_\mathrm{ss}=\min_{\vec\omega}\mathcal{M}_{\mathrm{ss}}(\vec\omega).
\end{equation}

Importantly, the obtainable accuracy is intimately related to the geometry of the linear functions we seek to measure. In particular, one finds the best performance for this strategy when $\mathcal{G}$ is approximately $d-1$; that is, when $\phi_{\vec{\omega},\ell}\approx 0$. This corresponds to the situations where the sensor symmetric states have the largest inter-sensor correlations $\mathcal{J}_\mathrm{opt}$ (i.e. are most entangled). We emphasize again, that there is no guarantee that this performance is always achievable, although in Ref.~\cite{rubio2020quantum} it was proven for $d=2$ and demonstrated for a large set of problems for $d>2$. 

\subsection{Naive Sequential Strategy}
In the naive sequential protocol, we sequentially measure the $n$ linear functions $\{f_1,\dots,f_n\}$ using an optimal single linear function protocol \cite{Eldredge2018}. For this, we determine the optimal times $t_\ell$ spent to measure the $\ell^{\mathrm{th}}$ function by minimizing Eq.~(\ref{eq:eoptorig}) for $C=I$ with respect to $\{t_1,\cdots, t_n\}$ under the constraint $\sum_\ell t_\ell=t$.
The solution to this Lagrange multiplier problem, presented in Appendix \ref{app:timealloc}, reads
\begin{equation}\label{eq:naivecost}
    \mathcal{M}_\mathrm{naive}=\frac{1}{t^2}\left(\sum_{\ell=1}^n [w_\ell\mu_\ell^2]^{1/3}\right)^3.
\end{equation}

As an important example, consider equal weights, $w_\ell \equiv \mathcal{N}/n$. Then we have \begin{equation}
    \frac{n^2\mathcal{N}}{dt^2}\leq\mathcal{M}_{\mathrm{naive}}\leq\frac{n^2\mathcal{N}}{t^2}.
\end{equation}
Indeed, the upper bound is obtained for unfavourable functions $\{f_\ell\}$ such that $\vec\mu=\vec{1}_n$ (``worst case"), with $\vec{1}_n$ the $n$-component vector of ones, whereas the lower bound is obtained for favourable functions $\{f_\ell\}$ with $\vec\mu=\vec{1}_n/\sqrt{d}$ (``best case"). These are the two extreme possible cases. Compared to the local protocol figure of merit of $\mathcal{N}/t^2$ for any choice of $w_\ell$, we see that in the worst case, the local protocol is always superior to the naive sequential protocol. Furthermore, even in the best case, we must have $d>n^2$ to obtain an advantage from the naive sequential protocol compared to the local protocol, implying a relatively large number of sensors. This shows that the naive sequential protocol, with $C=I$, is not very competitive. On the other hand, as we show now, by optimizing over $C$ a significant gain in accuracy over the local protocol can be achieved.

\subsection{Optimal Sequential Strategy}
Finally, we consider the optimal sequential protocol. The minimization over time in Eq.~(\ref{eq:eopt}) proceeds as in the naive case but with a general $C$. Therefore, again leaving details to Appendix \ref{app:timealloc}, we obtain for the optimal sequential protocol that 
\begin{align}\label{eq:eopt}
    \mathcal{M}_{\mathrm{opt}}=\min_C \frac{1}{t^2}\left[\sum_{\ell=1}^n\left(\sum_{m=1}^n w_m C_{m\ell}^2\right)^{\frac{1}{3}}{\mu'_\ell}^{2/3} \right]^3,
\end{align} 
with optimal time to measure the $\ell$th function given by 
\begin{align}\label{eq:timealloc}
    t_\ell=t\frac{\left(\sum_{m=1}^n w_mC_{m\ell}^2\right)^{1/3}{\mu'_\ell}^{2/3}}{\sum_{p=1}^n\left(\sum_{m=1}^n w_mC_{mp}^2\right)^{1/3}{\mu'_p}^{2/3}}.
\end{align}
Inserting the definition of $\mu_\ell'$ from Eq. (\ref{eq:mudef}), we arrive at
\begin{align}\label{eq:Eoptfull}
    &\mathcal{M}_{\mathrm{opt}}= \nonumber \\
    &\min_C \frac{1}{t^2}\left[\sum_{\ell=1}^n\left(\sum_{m=1}^n w_m C_{m\ell}^2\right)^{\frac{1}{3}}{\max_{i}\abs{\sum_{m=1}^n(C^{-1})_{\ell m}A_{m i}}}^{2/3} \right]^3.
\end{align}
Note that due to the appearance of both $C$ and $C^{-1}$ in the expression with the same powers, the result is invariant under a change in the normalization of the columns of $C$. Therefore we may fix these column normalizations and introduce the constraint that
\begin{equation}\label{eq:Cconstraint}
    \sum_{m=1}^n w_m C_{m\ell}^2=1,
\end{equation}
for each $\ell$. Under this constraint, we obtain the simpler expression
\begin{align}\label{eq:eoptconstr}
    \mathcal{M}_{\mathrm{opt}}=\min_C \frac{1}{t^2}\left[ \sum_{\ell=1}^n \max_{i}\abs{\sum_{m=1}^n(C^{-1})_{\ell m}A_{m i}}^{2/3} \right]^3,
\end{align} 
with optimal time per function given by 
\begin{align}\label{eq:timeallocconstr}
    t_\ell=t\frac{{\mu'_\ell}^{2/3}}{\sum_{m=1}^n{\mu'_m}^{2/3}}.
\end{align}

Geometrically, the constraint in Eq.~(\ref{eq:Cconstraint}) corresponds to restricting the columns of $C$ to the surface of an $(n-1)$-dimensional ellipsoid (or $(n-1)$-sphere if $w_m=\mathcal{N}/n$ $\forall \,m$). The columns of $C$ can then be efficiently parametrized by elliptical (or spherical) coordinates, and the optimization amounts to finding the best choice of corresponding angular variables. We emphasize that this choice of normalization can be made without loss of generality. 

We have now fully characterized our optimized sequential protocol. In particular, one can numerically perform the minimization over matrices $C$ in Eq.~(\ref{eq:Eoptfull}) subject to the constraint in Eq.~(\ref{eq:Cconstraint}). However, while for practical purposes we have solved the problem, many questions of more general nature arise at this point. In particular, what kind of advantage is provided by the optimized sequential protocol over the naive one? What geometries of coefficient vectors correspond to the best performance for the sequential protocol? How does it compare to the signed sensor symmetric approach? These questions will be addressed in the following section. All of the figures of merit calculated in this section are summarized in Table~\ref{tab:summaryoffom}.

\begin{table*}[]
    \centering
    \renewcommand{\arraystretch}{1.1} 
    \begin{tabular}{|c | c | c | c | c|}
    \hline
       & Local &  Naive Sequential & Signed Sensor Symmetric & Optimized Sequential  \\
    \hline
     \multicolumn{1}{|c|}{$\mathcal{M}$} & $\frac{\mathcal{N}}{t^2}$ & $\frac{1}{t^2}\left(\sum\limits_{\ell=1}^n [w_\ell\mu_\ell^2]^{1/3}\right)^3$  & $\begin{array}{c} \min\limits_\omega \frac{\mathcal{N}}{t^2}\frac{1+[d-2-\mathcal{G}(\vec\omega)]\mathcal{J}_\mathrm{opt}}{(1-\mathcal{J}_\mathrm{opt})[1+(d-1)\mathcal{J}_\mathrm{opt}]}\\  \mathcal{J}_\mathrm{opt}(\vec\omega)=\frac{1}{\mathcal{G}(\vec\omega)+2-d}\bigg[ 1-\sqrt{\frac{(\mathcal{G}(\vec\omega)+1)[d-1-\mathcal{G}(\vec\omega)]}{d-1}}\bigg] \\ \mathcal{G}(\vec\omega)=\frac{1}{\mathcal{N}}\sum\limits_{\ell=1}^n w_\ell\left(d\cos^2\phi_{\vec \omega, \ell}-1 \right) \end{array}$& $\begin{array}{c} \min\limits_C \frac{1}{t^2}\left[ \sum\limits_{\ell=1}^n \max\limits_{i}\abs{\sum\limits_{m=1}^n(C^{-1})_{\ell m}A_{m i}}^{2/3} \right]^3 \\ \text{subject to }  \sum\limits_{m=1}^n w_m C_{m\ell}^2=1 \end{array}$\\
     \hline
    \end{tabular}
    \caption{Summary of figures of merit. Recall, that for all strategies other than signed sensor symmetric strategy, we have an explicit physical protocol to achieve the given figure of merit. For the signed sensor symmetric strategy, beyond $d=2$, we are not necessarily guaranteed that a state exists that achieves the figure of merit, and therefore it is a lower bound, given the signed sensor symmetric state restriction.}
    \label{tab:summaryoffom}
\end{table*}

\section{Performance and Geometry}\label{s:compare}
To compare the performance of the different strategies, we first study some analytically accessible limits and then turn to a numerical analysis of the related optimization problem.

\subsection{Geometrically Symmetric Limit}\label{ss:geometricallysym}
\begin{figure}
    \centering
    \includegraphics[width = \columnwidth]{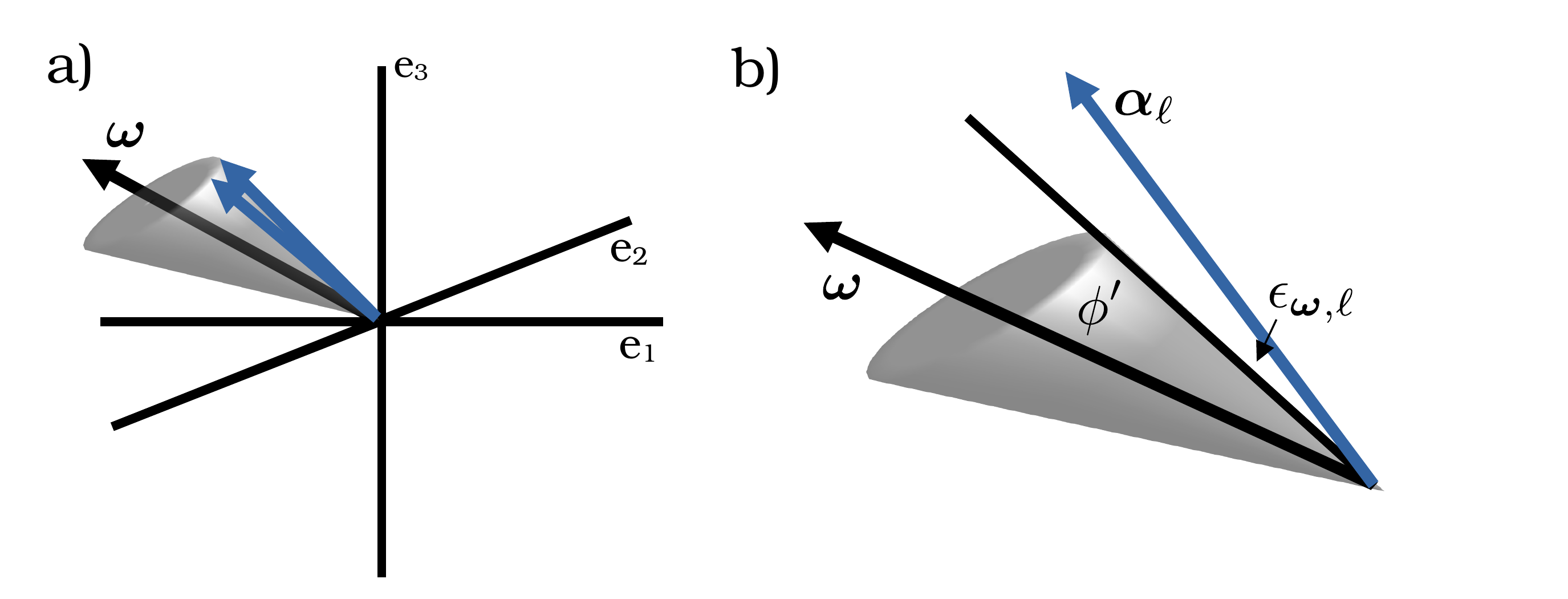}
    \caption{ (a) A visualization for $n=2$ functions and $d=3$ sensors of geometrically symmetric functions. In particular, the coefficent vectors lie near the surface of a cone centered on some $\vec\omega$. (b) The opening angle of the cone is given by $\phi'$ and the angular displacement from $\phi'$ for a particular $\vec\alpha_\ell$ is specified by $\epsilon_{\vec\omega,\ell}$, as defined in Eq.~(\ref{eq:epsdef}).}
    \label{fig:geometricallysym}
\end{figure}

We begin by considering what we refer to as the geometrically symmetric limit of the signed sensor symmetric strategy. This limit will be useful for comparing to the optimized sequential protocol in the following subsections. For this, we consider a situation where the coefficient vectors $\vec\alpha_\ell$ are all approximately the same angle $\phi'$ from some $\vec\omega$, which we recall is a vector with all elements $\pm 1$. This results in a particularly useful simplification of the expression for the geometry parameter $\mathcal{G}$. We then define
the parameter
\begin{equation}\label{eq:epsdef}
    \epsilon_{\vec\omega,\ell}=\phi_{\vec\omega,\ell}-\phi',
\end{equation}
so that $\epsilon_{\vec\omega,\ell}$ may be treated as a small parameter for a perturbative expansion, see Fig.~\ref{fig:geometricallysym}

The geometry parameter of the signed sensor symmetric strategy then reads
\begin{align}
\nonumber   \mathcal{G}(\vec\omega)&= \mathcal{G}_{\phi'}(\vec\omega)\\
\nonumber    &+\frac{1}{\mathcal{N}}\sum_{\ell=1}^n w_\ell d\Bigl(-2\epsilon_{\vec\omega,\ell}\sin\phi'\cos\phi'\nonumber-\epsilon_{\vec\omega,\ell}^2\cos(2\phi')\Bigr) \nonumber\\
 \label{eq:goverlaplimit}   &+\mathcal{O}\left(\epsilon_{\vec\omega,\ell}^3\right).
\end{align}
Here we expand in powers of $\epsilon_{\vec\omega,\ell}$ and define
\begin{align}
    \mathcal{G}_{\phi'}(\vec\omega) = \frac{1}{\mathcal{N}}\sum_{\ell=1}^n w_\ell\left(d\cos^2\phi'-1 \right)=d\cos^2\phi'-1,
\end{align}
the geometry parameter for measuring a single function at an angle $\phi'$ from $\vec\omega$. The condition on how small $\epsilon_{\vec\omega,\ell}$ needs to be depends on $\phi'$, but for any particular problem we can determine the necessary condition. In general, as long as $\epsilon_{\vec\omega,\ell}\ll 1/\sqrt{d}$, the corrections will be negligible.

Next we consider \cref{eq:sscost} in the large-$d$ limit and obtain 
\begin{align}\label{eq:largedlimit}
    \mathcal{M}_\mathrm{ss}&=\frac{\mathcal{N}}{t^2}\left(1-\frac{\mathcal{G}(\vec\omega)}{d}\right)\nonumber\\
    &\qquad+\mathcal{O}\left(\frac{\mathcal{N}}{dt^2}\sqrt{\frac{(1+\mathcal{G}(\vec\omega))(d-\mathcal{G}(\vec\omega)-1)}{d-1}}\right),
\end{align}
for arbitrary values of $\vec{\omega}$. We substitute \cref{eq:goverlaplimit} and obtain, to leading order in the geometrically symmetric limit and for large $d$, that
\begin{equation}\label{eq:larged}
    \mathcal{M}_\mathrm{ss}(\vec\omega)\approx \frac{\mathcal{N}}{t^2}\left(1-\frac{\mathcal{G}_{\phi'}(\vec\omega)}{d}\right)\approx\frac{\mathcal{N}}{t^2}\left(\sin^2\phi'+\frac{1}{d}\right).
\end{equation}
Note that, for $\phi'=0$, i.e. when all functions are nearly aligned with $\vec\omega$, this reduces to the expected optimal scaling $\mathcal{N}/(t^2d)$.  

We will use these results in the following sections as we compare the signed sensor symmetric strategy to the optimized sequential strategy.

\subsection{Nearly Overlapping Functions}
Next consider the case when all the vectors $\vec \alpha_\ell$ are ``close" in each component, i.e. we consider measuring a set of $n$ nearly identical functions. Intuitively, one would expect the optimal sequential strategy in this case to be spending almost all the time measuring the linear combination pointing towards the average of these functions, and then spending a small amount of time measuring in other directions in order to distinguish the small variations in the functions. We find that this intuition is rigorously true. We also find that, in this case, we can analytically determine a scaling advantage (in $d$) for this protocol relative to the signed sensor symmetric strategy (and, of course, the unentangled strategy). Finally, we consider a particular example from Ref.~\cite{rubio2020quantum} and find that its implication about the role of entanglement in protocol performance---namely that it can be disadvantageous in certain circumstances---is limited to the consideration of just the (unsigned) sensor symmetric strategy and is not generally true.

To formally define what we mean by ``nearly overlapping'', consider angles $\delta_{\ell}$ associated with each vector of coefficients $\vec\alpha_\ell$ as specified by
\begin{equation}
    \cos\delta_{\ell}=\vec\alpha_\ell\cdot\vec{\bar{a}},
\end{equation}
where $\vec{\bar a}$ is a vector, with Euclidean norm equal to 1, chosen such that the average angle $n^{-1}\sum_{\ell=1}^n \delta_\ell$ is minimized. For $\delta_\ell$ sufficiently small for all $\ell$, $\vec\alpha_\ell\approx\vec{\bar a}$ for all $\ell$. Furthermore,
\begin{equation}
    \max_{i}A_{\ell i} = \max_{i} {\bar a}_i + \mathcal{O}(\delta_\ell),
\end{equation}
for $A_{\ell i}=(\vec{\alpha}_\ell)_i$. Therefore, with $\delta=\max_\ell \delta_\ell$, we obtain from \cref{eq:eoptconstr} that
\begin{align}\label{eq:overlapstart}
    \mathcal{M}_\mathrm{opt}=&\frac{\max_i \bar{a}_i^2+\mathcal{O}(\delta^2)}{t^2}\min_C \left[\sum_{\ell=1}^n{\abs{\sum_{m=1}^n(C^{-1})_{\ell m}}}^{2/3} \right]^3.
\end{align}
Leaving the somewhat tedious details to \cref{app:overlappingfunctions}, we find that this reduces to the expected result that
\begin{align}\label{eq:nearlyoverlappingfinal}
    \mathcal{M}_\mathrm{opt}=\frac{\mathcal{N}}{t^2}\max_i \bar{a}_i^2+\mathcal{O}\left(\frac{\mathcal{N}\delta^2}{t^2}\right).
\end{align}
Note that, in general, $\delta\ll 1/\sqrt{d}$ ensures that this is a good leading-order approximation. This is a reduction in the variance by a factor of approximately (to order $\delta^2$) $\max_i \bar{a}_i^2\in [1/d,1]$ compared to the local protocol in \cref{eq:elocal}, or, when compared to the naive sequential protocol in \cref{eq:naivecost}, a reduction in the variance by a factor of order $\mathcal{O}(1/n^2)$. 

To compare to the signed sensor symmetric protocol, we note that this nearly overlapping case is merely a special case of the nearly geometrically symmetric case of the sensor symmetric protocol (provided $\delta$ is sufficiently small). In particular, $\delta$ is the relevant expansion parameter for our asymptotic approximations as $\epsilon_{\vec\omega,\ell}\leq \delta$ for all $\ell$. Therefore, to compare, we may simply use the previous results from \cref{ss:geometricallysym} with corrections upper bounded by taking $\epsilon_{\vec\omega,\ell}\rightarrow \delta$.

Furthermore, we note that, to leading order,  $\mathcal{M}_{\mathrm{ss}}=\mathcal{N}\mathcal{M}_{\mathrm{ss}}^{(n=1)}$, and similarly, \cref{eq:nearlyoverlappingfinal} also has the leading-order expression $\mathcal{M}_{\mathrm{opt}}=\mathcal{N}\mathcal{M}_{\mathrm{opt}}^{(n=1)}$, where the right-hand sides correspond to the accuracy $\mathcal{N}$ times the single-function estimation figure of merit. Therefore, we see that, in order to compare the accuracy of both protocols for nearly overlapping functions, it is sufficient to compare their performance for single-function estimation. 

Of course, for a single function, the ``sequential'' strategy is provably optimal as we have reduced it to the case of Ref.~\cite{Eldredge2018}. So, at best, the signed sensor symmetric strategy will perform the same as the ``sequential'' strategy for a single function. For example, we note that for the best case for both strategies---where all functions are oriented along some $\vec\omega$ to order $\mathcal{O}(\delta)$---both approaches have a cost to leading order of $\mathcal{N}/(t^2d)$, which is superior to the local protocol by $1/d$. Also, for $d=2$, the time-independent protocol of Ref. \cite{Eldredge2018} does actually utilize sensor symmetric states, because the initial states are chosen from the set
\begin{align}
    \ket\psi&=\frac{1}{\sqrt 2}\left(\ket{00}+\ket{11} \right)\nonumber\\
    \ket\psi&=\frac{1}{\sqrt 2}\left(\ket{01}+\ket{10} \right),
\end{align}
and therefore, for all choices of functions with $d=2$ (where both approaches provide explicitly saturable bounds), the two protocols are identical and optimal.

For $d>2$, on the other hand, as previously discussed, there may not exist physical states that obtain the figure of merit provided by the signed sensor symmetric strategy. However, even if we assume the figure of merit $\mathcal{M}_\mathrm{ss}$ is attainable, we shall see that the optimized sequential strategy can often be the superior choice. In this context, we consider two examples. First, we demonstrate a scaling advantage in $d$ for the sequential protocol in this nearly overlapping limit. Then we revisit the example from Eq. (38) of Ref.~\cite{rubio2020quantum} and demonstrate that the implication made that entanglement can be detrimental is an artifact of the (unsigned) sensor symmetric approach and that for the better performing sequential protocol, as well as the more general signed sensor symmetric approach, entanglement is useful.

\emph{Example 1:} To demonstrate an example of a scaling advantage of the sequential protocol over the signed sensor symmetric strategy, suppose we have $n$ nearly overlapping functions such that $\delta\ll 1/\sqrt{d}$ relative to the vector of coefficients given by 
\begin{equation}\label{eq:ex1}
    \vec{\bar a}=\frac{1}{\sqrt{(x^2-y^2)\kappa+y^2d}}\big(\underbrace{x, \cdots, x}_{\kappa}, \underbrace{y, \cdots, y}_{d-\kappa} \big)^T,
\end{equation}
where the first $\kappa$ elements are (up to normalization) $x\in\mathbb{R}$ and the last $d-\kappa$ elements are $y\in\mathbb{R}$. We assume $x,y=\mathcal{O}(1)$ and $\kappa=\mathcal{O}(d^\beta)$ for $\beta\in[0,1)$. Without loss of generality, suppose $x>y$. In this case, the cost of the optimized sequential strategy is straightforwardly obtained from \cref{eq:nearlyoverlappingfinal} to be
\begin{align}
    &\mathcal{M}_\mathrm{opt}=\frac{\mathcal{N}}{t^2}\left(\frac{x^2}{(x^2-y^2)\kappa+y^2d}\right)+\mathcal{O}\left(\frac{\mathcal{N}\delta^2}{t^2}\right)\nonumber \\
    &=\frac{\mathcal{N}}{t^2}\frac{x^2}{y^2d}\left(1-\frac{(x^2-y^2)\kappa}{y^2d} \right)+\mathcal{O}\left[\frac{\mathcal{N}}{t^2}\left(\delta^2+d^{2(\beta-1)}\right)\right],
\end{align}
where the second line comes from expanding in powers of $\kappa/d$.
For the signed sensor symmetric strategy for the same problem, we pick $\vec\omega$ such that $\vec\omega_i=\mathrm{sgn}(\bar{a}_i)$, which minimizes the angle between $\vec{\bar{a}}$ and $\vec\omega$. In the large $d$ limit, we may then use \cref{eq:larged} with
\begin{align}\label{eq:cos}
&\cos^2\phi'=\frac{(\vec{\bar{a}}\cdot\vec\omega)^2}{d}=\frac{\big[(|x|-|y|)\kappa+|y|d\big]^2}{d\big[(x^2-y^2)\kappa+y^2d\big]}.
\end{align}
We can perform an expansion of the numerator of \cref{eq:cos} in powers of $\kappa/d$ as
\begin{align}
    \big[(|x|&-|y|)\kappa+|y|d\big]^2=|y|^2d^2\left[1+\frac{(|x|-|y|)\kappa}{|y|d}\right]^2\nonumber\\
    &=|y|^2d^2\left[1+\frac{2(|x|-|y|)\kappa}{|y|d}+\mathcal{O}\left(\frac{\kappa^2}{d^2}\right)\right],
\end{align}
and expand the denominator as
\begin{align}
    &\frac{1}{d\big[(x^2-y^2)\kappa+y^2d\big]}=\frac{1}{y^2d^2}\left[1+\frac{(x^2-y^2)\kappa}{y^2d}\right]^{-1} \nonumber\\
    &\qquad=\frac{1}{y^2d^2}\left[1-\frac{(x^2-y^2)\kappa}{y^2d}+\mathcal{O}\left(\frac{\kappa^2}{d^2}\right)\right].
\end{align}
We then have
\begin{align}
    \sin^2\phi' = 1-\cos^2\phi' = \frac{(|x|-|y|)^2\kappa}{y^2 d}+\mathcal{O}\left(\frac{\kappa^2}{d^2}\right),
\end{align}
which we may plug into \cref{eq:larged} for the signed sensor symmetric strategy
\begin{equation}
     \mathcal{M}_\mathrm{ss}=\frac{\mathcal{N}}{t^2}\frac{(|x|-|y|)^2\kappa}{y^2 d}+\mathcal{O}\left[\frac{\mathcal{N}}{t^2}\left(\delta^2+d^{2(\beta-1)}\right)\right],
\end{equation}
which demonstrates a scaling advantage by a factor of $\mathcal{O}\left(\kappa^{-1}\right)=\mathcal{O}\left(d^{-\beta}\right)$ for the optimized sequential protocol in this problem.

\emph{Example 2:} Now we consider the example of a single function from Eq.~(38) of Ref.~\cite{rubio2020quantum} for $d=3$ sensors and coefficient vector \footnote{We have normalized differently by a factor of $1/\sqrt{3}$ from Eq.~(38) of Ref.~\cite{rubio2020quantum} in order to match our assumption that $||\vec\alpha||^2=1$.}
\begin{equation}
   \label{alphaExam}  \vec{\alpha}=\frac{1}{\sqrt{18}}\begin{pmatrix}
    \sqrt{2}+\sqrt{3}+1\\
     \sqrt{2}-\sqrt{3}+1\\
      \sqrt{2}-2
    \end{pmatrix}.
\end{equation}
The example was chosen in Ref.~\cite{rubio2020quantum} such that for $\vec\omega=\vec{1}$, $\mathcal{G}(\vec \omega)=0$, and thus $\mathcal{J}_{\rm opt}(\vec \omega)=0$, which in turn implies that the optimal (unsigned) sensor symmetric state is unentangled. Equation (\ref{eq:sscost}) then implies
\begin{equation}
    \mathcal{M}_{\mathrm{ss}}(\vec\omega=\vec{1})=\frac{1}{t^2},
\end{equation}
which is larger than the true optimal figure of merit, which is obtained by the ``sequential'' protocol:
\begin{equation}
     \mathcal{M}_{\mathrm{opt}}=\frac{1}{t^2}\left(\frac{\sqrt{2}+\sqrt{3}+1}{\sqrt{18}}\right)^2\approx \frac{0.9551}{t^2}.
\end{equation}
We also note  that, even within the framework of sensor symmetric strategies, the result obtained from Ref.~\cite{rubio2020quantum} is not the best one can do. If we extend to the signed sensor symmetric approach, one can consider $\vec\omega=(1,1,-1)^T$ and do better. In particular, in this case, one obtains
\begin{equation}
     \mathcal{M}_{\mathrm{ss}}(\vec\omega)=\frac{0.9554}{t^2},
\end{equation}
which is only slightly worse than the true optimum, and, crucially, also involves entanglement. Therefore, from this example, we learn that (a) entanglement \emph{is} helpful for measuring the function in Eq.~(\ref{alphaExam}), just not when we restrict to (unsigned) sensor symmetric states, and (b) accuracy is (unsurprisingly) potentially decreased when restricting ourselves to sensor symmetric states. 

For convenience, we summarize the analytic results comparing the signed sensor symmetric and optimized sequential strategies in \cref{tab:summary}.

\begin{table*}[]
    \centering
    \renewcommand{\arraystretch}{1.1} 
    \begin{tabular}{| c | c | c|}
    \hline
       Setting & Signed Sensor Symmetric & Optimized Sequential  \\
    \hline
     \multicolumn{1}{|c|}{$\begin{array}{c} \text{Geometrically symmetric limit}\\ (\text{large } d)\end{array}$} & $\begin{array}{c} \mathcal{M}_\mathrm{ss}(\vec\omega)\approx \frac{\mathcal{N}}{t^2}\left(1-\frac{\mathcal{G}_{\phi'}(\vec\omega)}{d}\right)\approx\frac{\mathcal{N}}{t^2}\left(\sin^2\phi'+\frac{1}{d}\right)\\ \phi':=\text{ angle of functions w.r.t. } \vec\omega\end{array}$ &     \\
    \hline
     \multicolumn{1}{|c|}{Nearly overlapping limit} & $\begin{array}{c} \text{Same as geometrically symmetric limit} \end{array}$& $ \begin{array}{c} \mathcal{M}_\mathrm{opt}=\frac{\mathcal{N}}{t^2}\max_i \bar{a}_i^2+\mathcal{O}\left(\frac{\mathcal{N}\delta^2}{t^2}\right) \\ \text{Functions aligned along } \vec{\bar{a}} \end{array}$\\
    \hline
    \multicolumn{1}{|c|}{$\begin{array}{c} \text{Best Case}\\\text{Functions aligned along some } \vec\omega \end{array}$} & $\begin{array}{c} \mathcal{M}_\mathrm{ss}=\frac{\mathcal{N}}{dt^2}\end{array}$ & $\begin{array}{c}  \mathcal{M}_\mathrm{opt}=\frac{\mathcal{N}}{dt^2}\end{array}$\\
    \hline
     \multicolumn{1}{|c|}{$\begin{array}{c} \text{Example 1 (Scaling)} \\ \text{Scaling advantage for } \mathcal{M}_\mathrm{opt} \\ \text{Functions aligned along \cref{eq:ex1}} \end{array}$} & $  \mathcal{M}_\mathrm{ss}=\mathcal{O}\left(\frac{\mathcal{N}\kappa}{dt^2}\right)=\mathcal{O}\left(\frac{\mathcal{N}}{d^{1-\beta}t^2}\right)\quad \left(\text{note: }\beta\in[0,1)\right)$ & $\mathcal{M}_\mathrm{opt}=\mathcal{O}\left(\frac{\mathcal{N}}{dt^2}\right)$\\
    \hline
    \end{tabular}
    \caption{Summary of analytic results comparing the signed sensor symmetric strategy and optimized sequential strategy.}
    \label{tab:summary}
\end{table*}

\subsection{Numerical Results}
\begin{figure*}
    \centering
    \includegraphics[width = \textwidth]{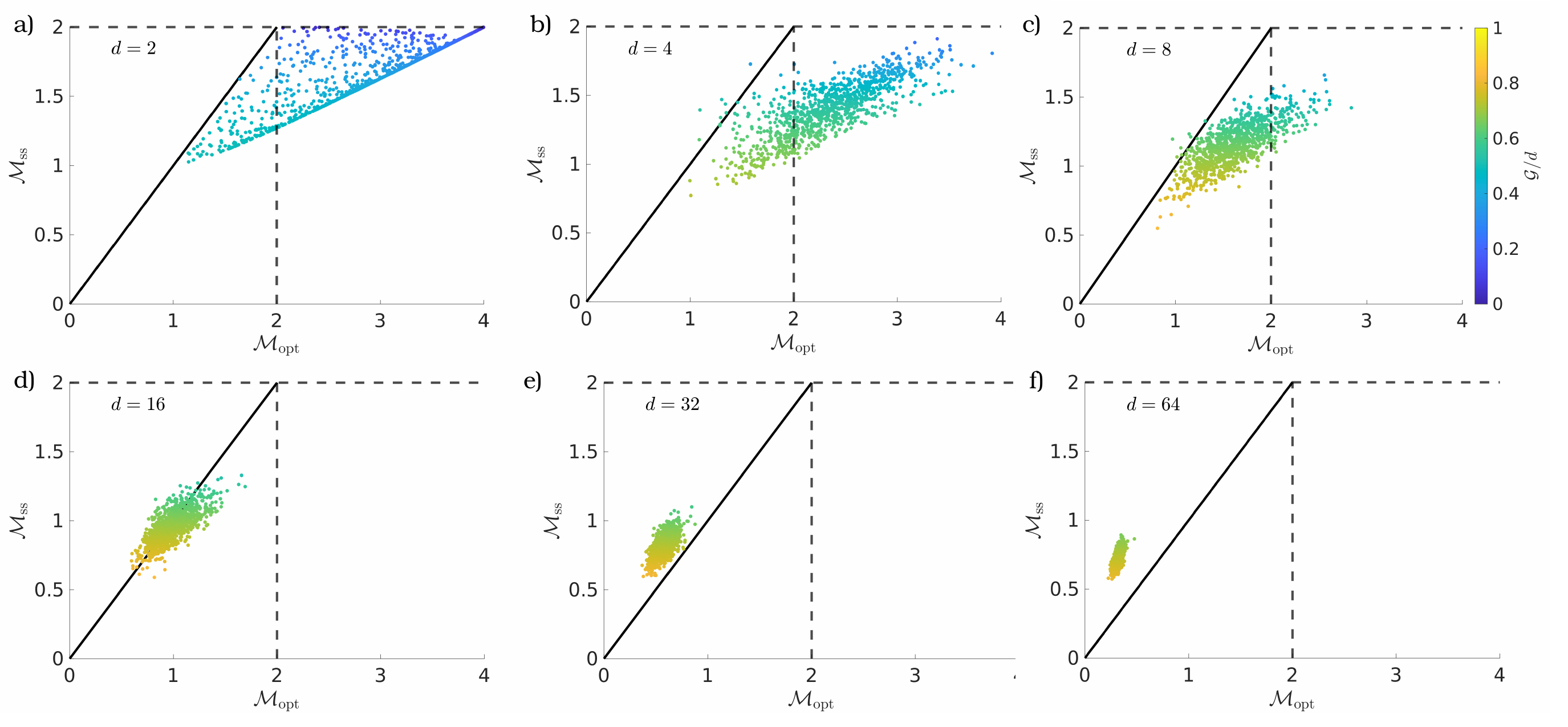}
    \caption{$\mathcal{M}_\mathrm{ss}$ versus $\mathcal{M}_\mathrm{opt}$ for 1000 random samples from the positive orthant of $\vec\alpha_1, \vec\alpha_2$ with $n=2$, $w_1=w_2=1$ for different numbers of sensors $d$. Dashed lines correspond to $\mathcal{M}_\mathrm{local}$. Colors correspond to the geometry parameter for the problem instance. Observe that the signed sensor symmetric approach is never worse than the local protocol, whereas the optimized sequential protocol can be. However, as $d$ increases the optimized sequential protocol is almost always superior. Also recall, that for $d>2$, $\mathcal{M}_\mathrm{ss}$ is generically just a lower bound, and it is not guaranteed one can achieve this figure of merit with physical states. Therefore, one can think of $\mathcal{M}_\mathrm{ss}$ as a best case scenario for a physically realized signed sensor symmetric protocol.}
    \label{fig:numerics}
\end{figure*}

In the previous sections, we found that both the optimized sequential and signed sensor-symmetric strategies perform identically (and optimally) when measuring many functions whose coefficient vectors $\{\vec\alpha_\ell\}$ are aligned along a particular $\vec\omega$. More generally, the optimized sequential protocol always performs at least as well as, and typically outperforms the signed sensor symmetric strategy when measuring many functions with nearly overlapping coefficient vectors, and in fact, we can obtain a scaling advantage in $d$ for certain problems (Example 1). However, while informative, the nearly overlapping limit considered above is such that the optimized sequential strategy performs its best. Therefore, it is of interest to also consider a broader class of examples and to consider where the signed sensor symmetric strategy outperforms the optimized sequential strategy. 

Unfortunately, however, a full analytic comparison between the different approaches is beyond reach as far as we know, so for a general problem, one must therefore compare the two approaches explicitly to see which one is the correct choice for a given situation. Here, to better understand the expected performance in such cases, we turn to numerics on random problem instances. Our key result is to demonstrate that generically, for large $d$, many problems are best approached using our optimized sequential protocol as opposed to the sensor symmetric or local strategies.

Numerically, the optimization over $C$ in Eq.~(\ref{eq:eoptconstr}), subject to Eq.~(\ref{eq:Cconstraint}), to obtain the cost of the optimized sequential protocol can be fairly costly in terms of computation time, as the optimization is non-convex and in a high dimensional parameter space. This is not necessarily an issue for particular applications, where only a limited number of such optimizations must be performed. As an example, consider $n=2$ functions, $d\geq n$ sensors, and equal weights in the figure of merit ($w_1=w_2=1$). The normalization condition (\ref{eq:Cconstraint}) implies that the columns of the $2\times 2$ matrix $C$ have unit length. We can parametrize this by two angles via
\begin{align}
  \label{EqC}  C = \begin{pmatrix} \cos \varphi & \cos \varphi' \\ \sin \varphi & \sin \varphi' \end{pmatrix}.
\end{align}
Given the coefficient vectors $\vec{\alpha}_{1,2}$ of the two functions to be estimated, the numerical optimization over $\varphi,\varphi'$ is accomplished straightforwardly. For $n=3$ functions, six angles $\varphi_1,\dots,\varphi_6$ are needed, making the optimization more challenging for larger $n$. 

The two functions, represented by the two normalized coefficient vectors $\vec{\alpha}_{1,2}$, depend on $2(d-1)$ real parameters. In this context, we randomly sample coefficients for the two functions from a uniform distribution and calculate the cost of the signed sensor symmetric strategy and the optimized sequential strategy. For $d=2^k$ for $k\in [1,6]$, we consider 1000 such problems where for simplicity we assume that $\vec\alpha_{1,2}$ are sampled from the positive orthant so that the optimal $\vec\omega$ is necessarily $\vec 1$ and plot the results in Fig.~\ref{fig:numerics}. 

\begin{figure}
    \centering
    \includegraphics[width=\columnwidth]{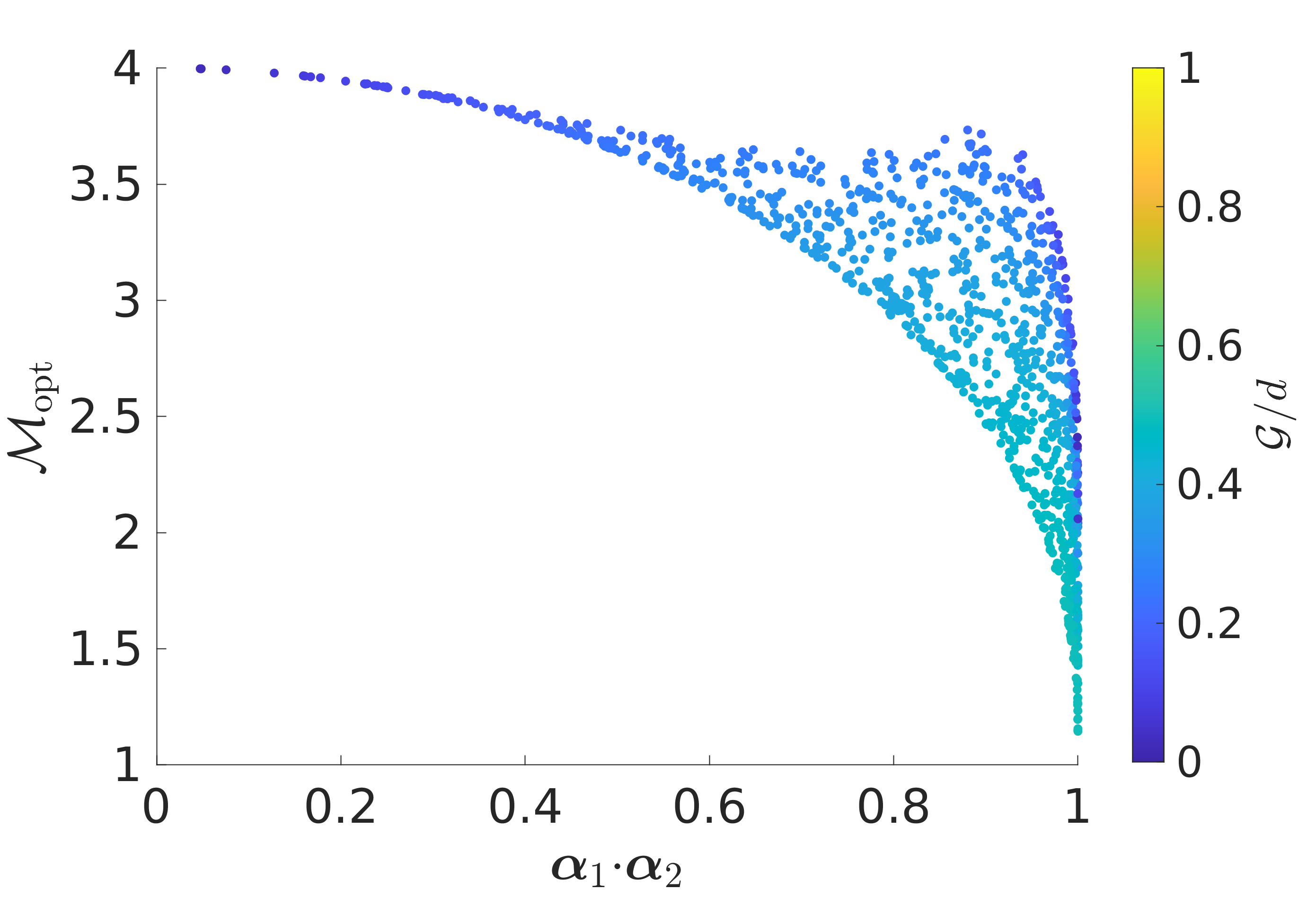}
    \caption{$\mathcal{M}_\mathrm{opt}$ versus $\vec\alpha_1\cdot\vec\alpha_2$ for $n=2$ functions and $d=2$ sensors. Note that the nearly orthogonal case ($\vec\alpha_1\cdot\vec\alpha_2\approx 0$) implies $\mathcal{M}_\mathrm{opt}\approx 4$ (i.e., the worst case for the optimized sequential strategy). Comparing to the first panel of \cref{fig:numerics} we see that in this case $\mathcal{M}_\mathrm{ss}\approx\mathcal{M}_\mathrm{local}=2$. \label{fig:d2}}
\end{figure}

We observe that the signed sensor symmetric strategy is never worse than the local protocol, whereas the optimized sequential protocol can be at small $d$. In the particular case of $n=d=2$,  the sequential strategy is never better than the signed sensor symmetric strategy. As previously mentioned, it is well known that,  for this problem,  when the two functions are orthogonal, a local protocol obtains the optimal variance (that is, $\mathcal{M}=\mathcal{N}/t^2$ is optimal) \cite{proctor2017networked, altenburg2018multi}. In this case, as demonstrated in Ref.~\cite{rubio2020quantum}, the sensor symmetric strategy matches this known optimal result. In particular, the sensor symmetric strategy predicts an optimal geometry parameter $\mathcal{G}(\omega)=0$, corresponding to no inter-sensor correlations and, therefore, a local protocol. We observe this behavior in panel a) of Fig.~\ref{fig:numerics} where the $\mathcal{G}=0$ points correspond to $\mathcal{M}_\mathrm{ss}=\mathcal{M}_\mathrm{local}=2$. Note that cases of $\mathcal{G}\approx 0$ that correspond to nearly orthogonal coefficient vectors are only those points where $\mathcal{M}_\mathrm{opt}\approx 4$, as can be concluded from Fig.~\ref{fig:d2} where we plot $\mathcal{M}_\mathrm{opt}$ versus $\vec{\alpha_1}\cdot\vec{\alpha_2}$. As $d$ increases, however, the optimized sequential protocol is almost always superior to both the local and signed sensor symmetric strategies for these randomized problem instances. 

\section{Conclusion and Outlook}
In this work, we explored the potential of sequential protocols to measure multiple functions with quantum sensor networks. We highlighted both analytical and numerical aspects, and compared the protocol to a generalized version of the sensor symmetric bounds for the same problem from Ref.~\cite{rubio2020quantum}. We find that, when $d$ is large, the sequential protocol is typically superior for generic problem instances. The sequential strategy also has the advantage of having an explicit protocol to obtain its given performance, whereas beyond $d=2$, while shown to be saturable in certain cases \cite{rubio2020quantum}, the lower bound when restricted to signed sensor symmetric states is not guaranteed to always be attainable. However, for a particular problem, one should compare both strategies, as neither is always superior. 

Our results, together with those in Ref.~\cite{rubio2020quantum}, point to an intriguing interplay between the geometric configuration of the functions to be measured and the performance of various protocols. In particular, our optimized sequential protocol performs best with nearly overlapping functions; the signed sensor symmetric approach performs best when the set $\{\vec \alpha_\ell\}$ is nearly aligned along some $\vec\omega$.
Beyond carefully tuned examples, we note that for most problems where we seek to estimate a collection of analytic functions of local field amplitudes, our protocol is the best known choice, especially with more than a small number of sensors $d$.

Our sequential protocol could directly be extended to the case where the sensors are each coupled to correlated field amplitudes as in the recent work by some of the authors \cite{qian2020optimal}; that is, instead of considering independent field amplitudes $\theta_i$ coupled to the sensors, one could consider the case where $\vec\theta$ is specified by a known analytic parameterization by some set of $k\leq d$ parameters. 

Our sequential protocol could also be extended to other physical settings beyond qubit sensors---namely, for any quantum sensor network where one may measure a single linear combination of field amplitudes, one can apply our sequential approach. For example, a collection of $d$ Mach-Zehnder interferometers could replace the qubit sensors, where the role of the local fields is played by interferometer phases \cite{holland1993interferometric, kim1998influence, devi2009quantum, dinani2016quantum, ge2018, gessner2018sensitivity}. Here, the limiting resource is the number of photons $N$ available to distribute among interferometers as opposed to the total time $t$. In this context, it was conjectured in  Ref.~\cite{proctor2018multiparameter} that one could measure a single function with variance $\mathcal{M} = \frac{\norm{\vec w}_1^2}{N^2}$---this replaces Eq.~(\ref{EqEld}), and otherwise everything remains the same. However, there are subtleties in the case where the average number of photons is not known \cite{hyllus2010entanglement}, which we do not consider here. Another relevant setting is the measurement of linear combinations of field-quadrature displacements as considered using an entanglement-enhanced continuous-variable protocol in  Ref.~\cite{zhuang2018distributed}. A variation of this protocol was experimentally implemented in Ref.~\cite{guo2020distributed}. One could also consider a combination of these settings where some field amplitudes are coupled to qubits, some to Mach-Zehnder interferometers, and some to field-quadrature displacements.

While the importance of geometry is striking, the general question of the information-theoretic optimal strategy that minimizes the quantum Fisher information for this problem remains a pressing open question. Additionally, our results are asymptotic and ignore the potential effects of decoherence. Understanding the performance of the sequential protocol in the non-asymptotic regime (i.e., via Bayesian analysis as considered in Ref.~\cite{rubio2020quantum}) and under the effects of decoherence remains a question of great importance.
These limitations aside, our findings advance the understanding of measuring multiple functions with quantum sensor networks and provides an alternative protocol that practically performs better than previously considered schemes in many instances.

\section{Acknowledgments}

We thank Timothy Qian for useful discussions. I.B., P.N., P.B., and A.V.G.\ acknowledge funding by AFOSR MURI, AFOSR, DoE QSA, NSF PFCQC program, ARO MURI, DoE ASCR Accelerated Research in Quantum Computing program (award No.~DE-SC0020312), U.S.~Department of Energy Award No.~DE-SC0019449, and the DoE ASCR Quantum Testbed Pathfinder program (award No.~DE-SC0019040). J.B.~acknowledges support by the U.S.~Department of Energy, Office of Science, Office of Advanced Scientific Computing Research, Department of Energy Computational Science Graduate Fellowship (award No.~DE-SC0019323).

\begin{appendix}

\section{Derivation of Signed Sensor Symmetric Bound}\label{app:signedsensorsymmetric}
In this appendix, we demonstrate that the explicit calculation of the inverse of the quantum Fisher information in Ref. \cite{rubio2020quantum} for sensor symmetric states can be extended to the signed sensor symmetric states of Eq.~(\ref{eq:signedsensorstates}). The calculation largely follows that in that reference. 

Begin by defining the symmetric matrix $\Omega=\vec \omega \vec\omega^T$ for $\vec\omega$ a vector with all elements $\pm 1$, as defined in the main text. For example, 
\begin{equation}
 \Omega=\begin{pmatrix} 1 \\ -1 \\ 1\end{pmatrix}\begin{pmatrix} 1 & -1 & 1\end{pmatrix}=\begin{pmatrix}
 1 & -1 & 1 \\
 -1 & 1 & -1 \\
 1 & -1 & 1
 \end{pmatrix}.
\end{equation}

Now, given an orthonormal basis $\{\hat e_i\}_{i\in[1,d]}$ for the real space where our vectors of coefficients $\{\vec\alpha_i\}$ are defined, we can write, for pure signed, sensor symmetric states,
\begin{align}
    \mathcal{F}_Q(\vec\theta)&=\sum_{i,j=1}^d t^2\left(\langle \sigma^{z}_i\sigma^{z}_j\rangle - \langle \sigma^{z}_i \rangle\langle \sigma^{z}_j \rangle\right)\hat e_i\hat e_j^T \nonumber \\
    &=4\left(v\sum_{i=1}^d \hat e_i\hat e_i^T+c\sum_{i\neq j} \Omega_{ij} \hat e_i\hat e_j^T \right) \nonumber \\
    &=4\left[(v-c)I+c\Omega \right]=4 v \left[(1-\mathcal{J})I+\mathcal{J}\Omega \right],
\end{align}
where $ \mathcal{F}_Q(\vec\theta)$ is the quantum Fisher information with respect to the parameters $\vec\theta$ and $v$ and $c$ are defined as in Eqs.~(\ref{eq:ssstates}) and (\ref{eq:signedsensorstates}) of the main text. We note this is equivalent to Eq. (22) of Ref. \cite{rubio2020quantum} but with $\vec 1\vec 1^T \rightarrow \Omega$, where $\vec{1}$ is the vector of $d$ $(+1)$s. To invert $\mathcal{F}_Q(\vec\theta)$ and evaluate the quantum Cram\'{e}r-Rao bound, we need the Fisher information matrix to be positive definite---i.e. we require its eigenvalues to be strictly positive. The characteristic equation for the eigenvalues $\lambda$ of $\mathcal{F}_Q(\vec\theta)$ is
\begin{equation}
    \det\left[4v\left((1-\mathcal{J}-\lambda/4v)I+\mathcal{J}\Omega\right) \right]=0.
\end{equation}
We then use the determinant identity \cite{horn2012matrix}: $\det(X+\vec x\vec y^T)=(1+\vec x^TX^{-1}\vec y)\det(X)$, with $X=[4v(1-\mathcal{J})-\lambda]I$, $\vec x=4v\mathcal{J}\vec\omega$, and $\vec y = \vec\omega$. With a bit of algebra, we obtain
\begin{equation}
    \left[4v(1+(d-1)\mathcal{J})-\lambda \right]\left[4v(1-\mathcal{J})-\lambda \right]^{d-1}=0,
\end{equation}
which is identical to Eq.~(24) in \cite{rubio2020quantum}. Here, we used that $\omega^T \omega = d$. Therefore, the eigenvalues of $\mathcal{F}_Q(\vec\theta)$ are $4v[1+(d-1)\mathcal{J}]$ with multiplicity one and $4v(1-\mathcal{J})$ with multiplicity $d-1$.  If we insist that the eigenvalues are positive (so that $\mathcal{F}_Q(\vt)$ is invertible), we then have the condition on $\mathcal{J}$ that $\mathcal{J}\in\left (\frac{1}{1-d}, 1 \right)$.

The inverse of $\mathcal{F}_Q(\vec\theta)$ is given by
\begin{equation}\label{eq:invF}
    \mathcal{F}_Q^{-1}(\vec\theta)=\frac{[1+(d-1)\mathcal{J}]I-\mathcal{J}\Omega}{4v(1-\mathcal{J})[1+(d-1)\mathcal{J}]}.
\end{equation}

We can verify this by computing
\begin{align}
    \mathcal{F}_Q^{-1}\mathcal{F}_Q&=\frac{[1+(d-1)\mathcal{J}]I-\mathcal{J}\Omega}{4v(1-\mathcal{J})[1+(d-1)\mathcal{J}]}(4 v) \left[(1-\mathcal{J})I+\mathcal{J}\Omega \right] \nonumber \\
    &=I,
\end{align}
where we used that $\Omega^2=\vec\omega\vec\omega^T\vec\omega\vec\omega^T=\Omega d$. We then may evaluate the quantum Cram\'{e}r-Rao bound
\begin{equation}\label{eq:bnd1}
    \mathcal{M}\geq \sum_{\ell=1}^n w_\ell (\mathcal{F}_Q^{-1}(\vec g) )_{\ell\ell} =\sum_{\ell=1}^n w_\ell (A \mathcal{F}_Q^{-1}(\vec\theta) A^T)_{\ell\ell}.
\end{equation}

Plugging Eq.~(\ref{eq:invF}) into Eq.~(\ref{eq:bnd1}) and using $4v=t^2$ for our Hamiltonian from Eq.~(\ref{eq:vc}), we obtain
\begin{align}\label{eq:bndapp}
    \mathcal{M}&\geq\sum_{\ell=1}^n \frac{[1+(d-2)\mathcal{J}]w_\ell(AA^T)_{\ell\ell}-w_\ell\mathcal{J}[A(\Omega-I)A^T]_{\ell\ell}}{t^2(1-\mathcal{J})[1+(d-1)\mathcal{J}]}\nonumber \\
    &\qquad=\frac{[1+(d-2)\mathcal{J}]\mathcal{N}-\mathcal{N}\mathcal{J}\mathcal{G}(\vec\omega)}{t^2(1-\mathcal{J})[1+(d-1)\mathcal{J}]}\nonumber \\
    &\qquad = \frac{\mathcal{N}}{t^2}\frac{1+(d-2-\mathcal{G}(\vec\omega))\mathcal{J}}{(1-\mathcal{J})[1+(d-1)\mathcal{J}]},
\end{align}
where we introduced the generalized geometry parameter 
\begin{align}
     \mathcal{G}(\vec\omega)&=\frac{1}{\mathcal{N}}\sum_{\ell=1}^n w_\ell[A(\Omega-I)A^T]_{\ell\ell}\nonumber \\
     &=\frac{1}{\mathcal{N}}\sum_{\ell=1}^n w_\ell\left[(A\vec\omega\vec\omega^T A^T)_{\ell\ell}-1 \right] \nonumber \\
     &=\frac{1}{\mathcal{N}}\sum_{\ell=1}^n w_\ell\left[(\vec\alpha_\ell\cdot\vec\omega)^2-1 \right] \nonumber \\
     &=\frac{1}{\mathcal{N}}\sum_{\ell=1}^n w_\ell\left(d\cos^2\phi_{\vec \omega, \ell}-1 \right).
\end{align}
Here $\mathcal{N}$ is the normalization factor as introduced in Eq.~(\ref{eq:norm}) in the main text and $\phi_{\vec \omega, \ell}$ is the angle between the linear functions specified by $\vec{\alpha}_\ell$ and $\vec \omega$. Note that $\mathcal{G}(\vec\omega)\in[-1, d-1]$. As in Appendix C of Ref.~\cite{rubio2020quantum}, we can find the optimal $\mathcal{J}$ in Eq.~(\ref{eq:bndapp}), provided $\mathcal{J}\in\left (\frac{1}{1-d}, 1 \right)$, to be
\begin{align}\label{eq:Jopt_app}
    \mathcal{J}_\mathrm{opt}&(\vec\omega)=\nonumber \\
    &\frac{1}{\mathcal{G}(\vec\omega)+2-d}\left[ 1-\sqrt{\frac{(\mathcal{G}(\vec\omega)+1)(d-1-\mathcal{G}(\vec\omega))}{d-1}}\right].
\end{align}
The ultimate best bound is found using
\begin{equation}
    \mathcal{M}_\mathrm{ss}=\min_{\vec\omega}\mathcal{M}_{\mathrm{ss}}(\vec\omega).
\end{equation}

\section{Optimal Time Allocation}\label{app:timealloc}
In this appendix, we consider the problem of optimal time division amongst the $n$ measured functions. In particular, given some matrix $C$, we want to compute the optimal times $\{t_1,\dots,t_n\}$ in
\begin{equation}
\begin{split}
    \mathcal{M}(C)&=\min_{\{t_1,\cdots, t_n\}}\left[\sum_{\ell=1}^n\sum_{m=1}^n w_m C_{m\ell}^2\left(\frac{\mu'_\ell}{t_\ell}\right)^2 \right] \\
    &=\min_{\{t_1,\cdots, t_n\}}\left[\sum_{\ell=1}^n t_\ell^{-2}\sum_{m=1}^n k_{m\ell} \right],
\end{split}
\end{equation}
subject to the constrain $\sum_{\ell=1}^n t_\ell=t$. In the second line, we define $k_{m\ell}=w_m C_{m\ell}^2{\mu'_\ell}^{2}$. Introducing a Lagrange multiplier $\gamma_0$, we obtain the $n+1$ equations
\begin{subequations}
\begin{align}
    \sum_{\ell=1}^n t_\ell&=t, \label{eqn:lagrange1} \\ 
    -\frac{2}{t_\ell^3}\sum_{m=1}^nk_{m\ell}&=\gamma_0 \quad \forall\, \ell. \label{eqn:lagrange2}
\end{align}
\end{subequations}
Solving the latter equations for each $t_\ell$ and inserting the solution into the first equation yields
\begin{equation}
    t=\left(-\frac{2}{\gamma_0}\right)^{\frac{1}{3}}\sum_{\ell=1}^n\left(\sum_{m=1}^nk_{m \ell}\right)^{\frac{1}{3}}.
\end{equation}
We can rearrange this to find the Lagrange multiplier
\begin{equation}
    \gamma_0=-\frac{2}{t^3}\left[\sum_{\ell=1}^n\left(\sum_{m=1}^nk_{m \ell}\right)^{\frac{1}{3}}\right]^3.
\end{equation}
Together with Eq. (\ref{eqn:lagrange2}) this gives the optimal time allocation
\begin{align}\label{eq:timealloc}
    t_\ell&=t\frac{\left( \sum_{m=1}^nk_{m\ell} \right)^{\frac{1}{3}}}{\sum_{\ell=1}^n\left(\sum_{m=1}^nk_{m\ell}\right)^{\frac{1}{3}}}\\
    &=t\frac{\left(\sum_{m=1}^n w_mC_{m\ell}^2\right)^{1/3}{\mu'_\ell}^{2/3}}{\sum_{k=1}^n\left(\sum_{m=1}^n w_mC_{mk}^2\right)^{1/3}{\mu'_k}^{2/3}}
\end{align}
and the time optimized figure of merit
\begin{equation}
    \mathcal{M}(C)=\frac{1}{t^2}\left[\sum_{\ell=1}^n\left(\sum_{m=1}^nk_{m\ell}\right)^{\frac{1}{3}} \right]^3.
\end{equation}

For the naive sequential protocol, we have $C=\mathbb{1}$ and $\vec\mu'=\vec\mu$, so that
\begin{equation}
    \mathcal{M}_{\mathrm{naive}}=\frac{1}{t^2}\left(\sum_{\ell=1}^n w_\ell^{1/3}\mu_\ell^{2/3}\right)^3.
\end{equation}

\section{Nearly Overlapping Functions}\label{app:overlappingfunctions}
Here we derive Eq.~(\ref{eq:nearlyoverlappingfinal}) from Eq.~(\ref{eq:overlapstart}). For this, consider the minimization over $C$ in  \cref{eq:overlapstart}.
To bound the expression, first note that, for any integer $\ell\in [1,n]$, we have
\begin{align}
\nonumber    1=\sum_{m=1}^n \delta_{\ell m}&=\sum_{m=1}^n\sum_{p=1}^n C_{\ell p}(C^{-1})_{pm}\\
  \nonumber  &\leq\sum_{p=1}^n\abs{C_{\ell p}}\abs{\sum_m(C^{-1})_{pm}}\\
 \label{eq:ineqstart}   \implies 1&\leq\left(\sum_{p=1}^n\abs{C_{\ell p}}\abs{\sum_m(C^{-1})_{pm}}\right)^2.
\end{align}
This inequality is true for all $C$. Also note that
\begin{align}\label{eq:j*}
 \left(\sum_{p=1}^n\abs{C_{\ell p}}\abs{\sum_m(C^{-1})_{pm}}\right)^2 \geq \sum_{p=1}^n\abs{C_{\ell p}}^2\abs{\sum_m(C^{-1})_{pm}}^2.
\end{align}
This inequality is an equality when $\sum_m(C^{-1})_{pm}=0$ for all but one single $p=p^*$. When this condition is satisfied, we consequently have
\begin{equation}\label{eq:intermclose}
    1\leq \sum_{p=1}^n\abs{C_{\ell p}}^2\abs{\sum_m(C^{-1})_{pm}}^2.
\end{equation}
Now take a weighted sum over $\ell  $ in \cref{eq:intermclose} and obtain
\begin{align}
    \sum_{\ell  =1}^n w_\ell  \leq& \sum_{\ell  =1}^n w_\ell  \sum_{p=1}^n\abs{C_{\ell  p}}^2\abs{\sum_m(C^{-1})_{pm}}^2 \nonumber \\
    &=\sum_{p=1}^n\abs{\sum_m(C^{-1})_{pm}}^2,
\end{align}
where in the second line we used the normalization from Eq.~(\ref{eq:Cconstraint}). Next use subadditivity, $\sum_p |x_p| \leq (\sum_p |x_p|^{1/3})^3$, to obtain
\begin{align}
  \label{TheIneq}  \sum_{\ell  =1}^n w_\ell  \leq \left[\sum_{p=1}^n{\abs{\sum_{\ell=1}^n(C^{-1})_{p\ell}}}^{2/3} \right]^3
\end{align}
valid for all $C$. The expression on the right is the one we need to minimize (over $C$) in Eq.~(\ref{eq:overlapstart}). Consequently, if we can saturate the last inequality, we have found the minimum of the expression, and arrive at
\begin{align}\label{eq:nearlyoverlappingfinal2}
    \mathcal{M}_\mathrm{opt}=\frac{\max_m \bar{a}_m^2}{t^2}\sum_{\ell  =1}^n w_\ell  +\mathcal{O}\left(\frac{\mathcal{N}\delta^2}{t^2}\right)
\end{align}
for nearly overlapping functions. We can, in fact, saturate the inequality (\ref{TheIneq}). Recall that, in order to saturate Eq.~(\ref{eq:j*}), we require the existence of an index $p^*$ such that
\begin{equation}\label{eq:cond1}
    \sum_{\ell =1}^n (C^{-1})_{p\ell  }=0
\end{equation}
for $p\neq p^*$ and otherwise
\begin{equation}\label{eq:cond2}
    \left(\sum_{\ell  =1}^n C_{p^*\ell  }^{-1}\right)^2=\sum_{\ell  =1}^n w_\ell  .
\end{equation}
Furthermore, we must satisfy the normalization condition in Eq.~(\ref{eq:Cconstraint}) for each column of $C$. Geometrically, this normalization constraint forces each column of $C$ to be on the surface of an ellipsoid in $n$-dimensional space.

Suppose the row vector $C^{-1}_{p^*}=(\sqrt{\sum_{\ell  =1}^n w_\ell  }/n)\vec{1}^T$. This clearly satisfies Eq.~(\ref{eq:cond2}). We can satisfy Eq.~(\ref{eq:cond1}) by noting that Eq.~(\ref{eq:cond1}) can be written as $\vec{1}\cdot C_{p\neq p^*}^{-1}=0$. Therefore, Eq.~(\ref{eq:cond1}) is satisfied if the rows $p\neq p^*$ of $C^{-1}$ are orthogonal to $C^{-1}_{p^*}$---that is, they exist in $\vec{1}^\perp$.

It remains to show that we can choose such a $C$ that satisfies Eq.~(\ref{eq:Cconstraint}). We have then that the column vector of $C$, $C_{p^*}=(1/\sqrt{\sum_{\ell  =1}^n w_\ell  })\vec{1}^T$ which satisfies both $C_{p^*}\cdot C^{-1}_{p^*}=1$ and 
\begin{equation}
    \sum_{\ell  =1}^n w_\ell   C_{\ell   p^*}^2=1.
\end{equation}
The remaining columns of $C$ exist in $\vec{1}^\perp$ and must exist on the $n$-dimensional ellipsoid specified by Eq.~(\ref{eq:Cconstraint}). 

As $\vec{1}^\perp$ is a subspace geometrically represented as a hyperplane through the origin, it necessarily intersects the ellipsoid (centered on the origin) specified by Eq.~(\ref{eq:Cconstraint}) forming an ellipsoid of dimension $n-1$. Therefore, we can satisfy all constraints and saturate Eq.~(\ref{eq:j*}).

Furthermore, we can confirm this choice of $C$ also saturates Eq.~(\ref{eq:ineqstart}) as 
\begin{align}
    &\left(\sum_{p=1}^n\abs{C_{\ell p}}\abs{\sum_m(C^{-1})_{pm}}\right)^2= \abs{C_{\ell p^*}}^2\abs{\sum_m(C^{-1})_{p^*m}}^2 \nonumber \\
    &=\frac{1}{\sum_{\ell=1}^n w_\ell} \sum_{\ell=1}^n w_\ell =1,
\end{align}
and so we have confirmed we may obtain the equality as in \cref{eq:nearlyoverlappingfinal2}.

\end{appendix}

\end{document}